\documentclass[11pt,a4paper]{article}
\pdfoutput=1

\usepackage{jheppub}
\usepackage{latexsym}
\usepackage{revsymb}
\usepackage{multirow}
\usepackage{color}
\usepackage[usenames,dvipsnames,svgnames,table]{xcolor}

\usepackage{graphicx}
\usepackage{epsfig}  
\usepackage{epsf}    
\usepackage{dcolumn}
\usepackage{bm}
\usepackage{dcolumn}
\usepackage{textcomp}
\usepackage{float}
\usepackage{subfig}
\usepackage{hypcap}
\usepackage[]{hyperref}

\hypersetup{
  bookmarks=true,         
  unicode=false,          
  pdftoolbar=true,        
 pdfmenubar=true,        
 pdffitwindow=true,     
 pdfstartview={FitH},    
 pdfsubject={Neutrino Oscillations Phenomenology},   
 pdfnewwindow=true,      
 pdfcreator={RevTeX},
 colorlinks=true,       
 linkcolor=red,          
 citecolor=blue,        
filecolor=black,      
 urlcolor=blue,           
  }

\newcommand{\be}{\begin{equation}}
\newcommand{\ee}{\end{equation}}
\newcommand{\ba}{\begin{eqnarray}}
\newcommand{\ea}{\end{eqnarray}}
\def\nue{{\nu_e}}

\def\numu{{\nu_{\mu}}}

\def\anu{{\bar\nu}}

\newcommand{\beq}{\begin{equation}}
\newcommand{\eeq}{\end{equation}}
\newcommand{\beqa}{\begin{eqnarray}}
\newcommand{\eeqa}{\end{eqnarray}}

\newcommand{\ty}{{\theta_{13}}}
\newcommand{\tz}{{\theta_{23}}}

\newcommand{\sa}{\sin^2 \theta_{23}}

\newcommand{\ms}{\Delta m^2_{21}}
\newcommand{\ma}{\Delta m^2_{31}}
\newcommand{\dcp}{\delta_{\mathrm{CP}}}
\newcommand{\pmue}{P(\nu_\mu \rightarrow \nu_e)}

\newcommand{\dchsq}{\Delta\chi^2}

\newcommand{\ie}{{\it i.e.}}

\title{Exploring the three flavor effects with future superbeams using liquid argon detectors}

\author[a]{Sanjib Kumar Agarwalla,}
\author[b]{Suprabh Prakash,} 
{\author[b]{S. Uma Sankar$\,$}

\affiliation[a]{Institute of Physics, Sachivalaya Marg, Sainik School Post, Bhubaneswar 751005, India}
\affiliation[b]{Department of Physics, Indian Institute of Technology Bombay, Mumbai 400076, India}

\emailAdd{sanjib@iopb.res.in}
\emailAdd{suprabh@phy.iitb.ac.in}
\emailAdd{uma@phy.iitb.ac.in}

\abstract
{
Recent measurement of a moderately large value of $\theta_{13}$ signifies an important breakthrough in establishing
the standard three flavor oscillation picture of neutrinos. It has provided an opportunity to explore the sub-dominant 
three flavor effects in present and future long-baseline experiments. In this paper, we perform a comparative study of the 
physics reach of two future superbeam facilities, LBNE and LBNO in their first phases of run, to resolve the issues of
neutrino mass hierarchy, octant of $\tz$, and leptonic CP violation. We also find that the sensitivity of these future facilities 
can be improved significantly by adding the projected data from T2K and NO$\nu$A. Stand-alone LBNO setup with a 10 kt detector 
has a mass hierarchy discovery reach of more than 7$\sigma$, for the lowest allowed value of $\sa$(true) = 0.34.
This result is valid for any choice of true $\dcp$ and hierarchy. LBNE10, in combination with T2K and NO$\nu$A, 
can achieve 3$\sigma$ hierarchy discrimination for any choice of $\dcp$, $\sa$, and hierarchy.
The same combination can provide a 3$\sigma$ octant resolution for $\sa$(true) $\leq$ 0.44 or for $\sa$(true) $\geq$ 0.58 
for all values of $\dcp$(true). LBNO can give similar results with 10 kt detector mass. 
In their first phases, both LBNE10 and LBNO with 20 kt detector can establish leptonic CP violation for around 50\% values of 
true $\dcp$ at $2\sigma$ confidence level. In case of LBNE10, CP coverage at 3$\sigma$ can be enhanced from 3\% to 43\% by 
combining T2K and NO$\nu$A data, assuming $\sa$(true) = 0.5. For LBNO setup, CP violation discovery at 3$\sigma$ is possible 
for 46\% values of true $\dcp$ if we add the data from T2K and NO$\nu$A.
}

\keywords{Neutrino Mass Hierarchy, Octant of $\theta_{23}$, Leptonic CP violation, Long-baseline Experiments, Three Flavor Effects}
\arxivnumber{1304.3251}

\begin{document}
\maketitle
\flushbottom

\section{Introduction and Motivation}
\label{introduction}

The discovery of neutrino oscillations over the past decade provides firm evidence for new physics. 
Recently, the unknown 1-3 lepton mixing angle has been measured quite precisely
by the reactor experiments~\cite{An:2013kva,An:2012eh,Ahn:2012nd,Abe:2012tg}. 
They have found a moderately large value, not too far from its previous 
upper bound. This represents a significant milestone towards addressing 
the remaining fundamental questions, in particular determining the 
neutrino mass hierarchy and searching for CP violation in the neutrino sector. 
Another recent and crucial development is the indication of 
non-maximal 2-3 mixing by the MINOS accelerator 
experiment~\cite{Nichol:2013caa,Adamson:2013whj}, leading to the 
problem of determining the correct octant of $\tz$. 
It is possible to resolve all the above three issues by the observation
of $\nu_e$ appearance via $\nu_\mu \rightarrow \nu_e$ oscillations. 
The determination of CP violation in particular requires the full 
interplay of three flavor effects in neutrino oscillations.

Oscillation data are insensitive to the lowest neutrino mass. However, 
it can be measured in tritium beta decay processes~\cite{Osipowicz:2001sq}, 
neutrinoless double beta decay experiments~\cite{Avignone:2007fu}, 
and from the contribution of neutrinos to the energy density of the 
universe~\cite{Lesgourgues:2012uu}.
Very recent data from the Planck experiment in combination with the 
WMAP polarization and baryon acoustic oscillation measurements 
have set an upper bound on the sum of all the neutrino mass eigenvalues 
of $\sum m_i \leq 0.23$ eV at $95\%$ C.L.~\cite{Ade:2013zuv}.
But, oscillation experiments are capable of measuring the two independent 
mass-squared differences: $\ms = m_2^2 - m_1^2$ and 
$\ma = m_3^2 - m_1^2$. $\ms$ is required to be positive by the solar 
neutrino data but at present $\ma$ can be either positive or negative.
Hence, two patterns of neutrino masses are possible: $m_3 > m_2 > m_1$, 
called normal hierarchy (NH) where $\ma$ is positive and 
$m_2 > m_1 > m_3$, called inverted hierarchy (IH) where $\ma$ is negative. 
Leptonic CP violation can be established if CP violating phase
$\dcp$ in the mixing matrix, differs from both 0 and $180^\circ$. 
So far, there is no constraint on $\dcp$. It can take any value
in the range $[-180^\circ,180^\circ]$. Regarding $\tz$, 
all global fits~\cite{Tortola:2012te,Fogli:2012ua,GonzalezGarcia:2012sz} 
point to a deviation from maximal mixing (MM) \ie\,\,$(0.5 - \sa) \ne 0$. 
This raises an additional question: ``whether $\theta_{23}$ lies in the 
lower octant (LO: $\theta_{23}<45^\circ$) or higher octant 
(HO: $\theta_{23}>45^\circ$)?''. 

Settling the issue of neutrino mass hierarchy is crucial to determine 
the structure of neutrino mass matrix. This structure will provide the
fundamental input needed to develop the theory of neutrino masses and 
mixing~\cite{Albright:2006cw}. Neutrino mass hierarchy is also a key 
parameter for neutrinoless double beta decay searches probing the 
Majorana nature of neutrinos~\cite{Pascoli:2005zb}.
Another fundamental issue that needs to be addressed in long-baseline 
experiments is to establish leptonic CP violation and measure $\dcp$. 
This new CP violation in the lepton sector may be able  
to explain the observed matter anti-matter asymmetry in the universe 
via leptogenesis~\cite{Fukugita:1986hr}.
A number of innovative ideas, such as $\mu \leftrightarrow \tau$ 
symmetry~\cite{Mohapatra:1998ka}, $A_4$ flavor symmetry~\cite{Babu:2002dz}, 
quark-lepton complementarity~\cite{Minakata:2004xt}, and 
neutrino mixing anarchy~\cite{Hall:1999sn,deGouvea:2012ac} 
have been invoked to explain the observed
pattern of one small and two large mixing angles in the neutrino sector. 
Measurements of the precise values of $\ty$ and $\tz$ 
will reveal the pattern of deviations from these symmetries and will 
lead to a better understanding of neutrino masses and mixing. 
In particular, the resolution of $\tz$ octant will severely constrain 
the patterns of symmetry breaking.
With the recent discovery of moderately large value of $\theta_{13}$, 
these three fundamental measurements fall within our reach.

The combined data from the current $\nu_e$ appearance experiments, 
T2K~\cite{Itow:2001ee,Abe:2011ks} and 
NO$\nu$A~\cite{Ayres:2002ws,Ayres:2004js,Ayres:2007tu},
can provide a hint at $90\%$ confidence level for 
neutrino mass ordering~\cite{Agarwalla:2012bv} and at $95\%$ confidence 
level for octant of $\tz$~\cite{Agarwalla:2013ju,Chatterjee:2013qus}. 
They can determine these quantities at $> 99\%$ C.L. only for a very small 
range of favorable values of $\dcp$. Discovery of leptonic CP violation 
is possible at 95\% C.L. only for values of $\dcp$ close to $\pm 90^\circ$,
{\it i.e.} where CP violation is maximum~\cite{Agarwalla:2012bv}. Hence, 
future facilities consisting of intense, high power wide-band beams and 
large smart detectors are mandatory to cover the entire parameter space at 
a high confidence level. In this paper, we explore the capabilities of 
future superbeam experiments with liquid argon detectors, 
LBNE~\cite{Diwan:2003bp,Barger:2007yw,Huber:2010dx,Akiri:2011dv,Adams:2013qkq} and
LBNO~\cite{Autiero:2007zj,Rubbia:2010fm,Angus:2010sz,Rubbia:2010zz,Stahl:2012exa} 
towards resolving these unknowns. We first present the stand-alone 
performances of these setups in their first phases. Then we examine 
how the addition of projected data from T2K and NO$\nu$A, 
can improve the sensitivity of these future facilities.
We also study in detail how these sensitivities change as the true 
value of $\sa$ varies in its allowed 3$\sigma$ range of 0.34 to 0.67.


We start with a brief discussion of $\numu \rightarrow \nue$ oscillation 
channel in section~\ref{sec:three-flavor}. In section~\ref{experiments}, 
we describe the important features of the experimental setups
under consideration. Next, we introduce the concept of bi-events 
plots ($\nu_e$ vs. $\bar{\nu}_e$ appearance events) 
to explain the underlying physics in section~\ref{bi-events-plots}. 
In section~\ref{results}, we present our results. 
Finally, we summarize and draw our conclusions in section~\ref{conclusions}. 

\section{Platinum Channel: Test Bed for Three Flavor Effects }
\label{sec:three-flavor}

A study of $\nu_\mu \rightarrow \nu_e$ and $\bar{\nu}_\mu \rightarrow \bar{\nu}_e$ 
oscillations at long-baseline superbeam experiments 
is the simplest way to probe three flavor effects, including 
sub-leading ones. Such a study is capable of achieving all the three objectives 
mentioned in section~\ref{introduction}. An approximate analytic expression for 
the oscillation probability, $P_{\mu e}$, 
in matter~\cite{Wolfenstein:1977ue,Mikheev:1986gs,Barger:1980tf}, 
is given by

\ba
P_{\mu e} &\simeq&
{\underbrace{\sin^2\theta_{23} \sin^22\theta_{13}
    \frac{\sin^2[(1-\hat{A})\Delta]}{(1-\hat{A})^2}}_{C_0}}
    + {\underbrace{\alpha^2 \cos^2\theta_{23} \sin^22\theta_{12}
    \frac{\sin^2(\hat{A}\Delta)}{{\hat{A}}^2}}_{C_1}} \nonumber \\
&-& {\underbrace{\alpha \sin2\theta_{13}\cos\theta_{13} \sin2\theta_{12}
    \sin2\theta_{23} \sin(\Delta) \frac{\sin(\hat{A}\Delta)}{\hat{A}}
    \frac{\sin[(1-\hat{A})\Delta]}{(1-\hat{A})}}_{C_-}} \sin\dcp \nonumber \\
&+& {\underbrace{\alpha \sin2\theta_{13}\cos\theta_{13} \sin2\theta_{12}
    \sin2\theta_{23} \cos(\Delta) \frac{\sin(\hat{A}\Delta)}{\hat{A}}
    \frac{\sin[(1-\hat{A})\Delta]}{(1-\hat{A})}}_{C_+}} \cos\dcp ,
\label{eq:pmue}
\ea
where
\ba
\Delta\equiv \frac{\ma L}{4E},
~~
\hat{A} \equiv \frac{A}{\ma},
~~
 A=\pm 2\sqrt{2}G_FN_eE.
\label{eq:matt}
\ea
Equation~\ref{eq:pmue} has been derived under the constant matter 
density approximation, keeping terms only up to second order in the 
small quantities $\theta_{13}$ and 
$\alpha \equiv \ms/\ma$~\cite{Cervera:2000kp,Freund:2001ui,Akhmedov:2004ny}.
Here, $A$ is the matter potential, expressed in terms of the electron density, 
$N_e$, and the (anti-)neutrino energy $E$. It is positive for neutrinos and 
negative for anti-neutrinos. For anti-neutrinos, the term
proportional to $\sin \dcp$ has the opposite sign. 
So far, it was possible to analyze the
data from each oscillation experiment using an appropriate, effective
two flavor oscillation approach because of the smallness of the mixing
angle $\sin2\theta_{13} \simeq 0.3$ and the ratio $\alpha \simeq 0.03$. 
This method has been quite successful in measuring 
the solar and atmospheric neutrino parameters. The next step must 
involve probing the full three flavor effects, including the sub-leading 
ones proportional to $\alpha$. This task will be undertaken, 
for the first time, by the current generation experiments T2K and NO$\nu$A.

In this paper, we consider two future long-baseline superbeam 
experiments with large matter effect.
The matter effect increases $\pmue$ oscillation probability for NH 
and decreases it for IH. For anti-neutrinos the situation is reversed.
It can be seen from equation~\ref{eq:pmue} that the dominant term 
($C_0$) is driven by matter modified $\ma$ and is proportional 
to $\sin^2 \theta_{23}\sin^22\theta_{13}$ but the sub-dominant 
$\dcp$ dependent terms ($C_{-}$ \& $C_{+}$) are suppressed by $\alpha$.
Since the hierarchy and $\dcp$ are both unknown, the interplay of 
the terms $C_0$, $C_{-}$, and $C_{+}$ in equation~\ref{eq:pmue} 
gives rise to hierarchy-$\dcp$ degeneracy~\cite{Minakata:2001qm}. 
If the matter effects are large enough, this degeneracy can be broken 
completely. This is {\em not} the case for T2K and NO$\nu$A, because 
of which their sensitivity to hierarchy is modest for about half
the $\dcp$ range. There is a similar octant-$\dcp$ degeneracy also, 
which limits our ability to determine the correct octant of $\theta_{23}$.
This problem can be solved by having substantial data in both 
$\nu$ and $\bar{\nu}$ channels~\cite{Agarwalla:2013ju}.
Both the future facilities, LBNE (baseline of 1300 km) and 
LBNO (baseline of 2290 km) will operate at multi-GeV energies with very 
long-baselines. This will lead to a large enough matter effect which breaks 
the hierarchy-$\dcp$ degeneracy completely.
They are also scheduled to have equal $\nu$ and $\bar{\nu}$ runs, 
and can resolve the octant-$\dcp$ degeneracy effectively.
These experiments are planning to use liquid argon time projection 
chambers (LArTPCs) which have excellent kinematic reconstruction 
capability for all the observed particles. This feature helps in 
rejecting almost all of the large neutral current background.

\section{Experimental Specifications}
\label{experiments}

In this section, we briefly describe the key experimental features 
of the current (off-axis) and future (on-axis) generation 
long-baseline superbeam experiments that we use in our simulation.

\subsection{Current Generation: T2K and NO$\nu$A}
\label{current}

In Japan, the Tokai-to-Kamioka (T2K) experiment~\cite{Itow:2001ee,Abe:2011ks} 
started taking data in 2010.  
The NO$\nu$A experiment~\cite{Ayres:2002ws,Ayres:2004js,Ayres:2007tu} 
in the United States is now under construction 
and will start taking data near the end of this year. 
The main goal of these experiments is to detect the electron neutrino 
appearance events in a $\numu$ beam using the classic off-axis beam 
technique~\cite{Para:2001cu} that delivers a beam with a narrow 
peak in the energy spectrum. The position of this peak is tuned 
to be close to the expected oscillation maximum. 
In our study, we 
have explored the improvement in the physics capabilities of
LBNE and LBNO in their first phases, 
due to the addition of the 
projected data from T2K and NO$\nu$A experiments.

In the T2K experiment, a $2.5^\circ$ off-axis $\nu_\mu$ beam from J-PARC is observed in the 
Super-Kamiokande detector (fiducial volume 22.5 $\mathrm{kt}$) at Kamioka, at a distance of 
295 $\mathrm{km}$~\cite{Itow:2001ee}.
The neutrino flux peaks sharply at the first oscillation maximum of 
$0.6$ $\mathrm{GeV}$. For mass hierarchy and 
CP violation studies, we consider 5 years of neutrino run with a 
beam power of $0.75$ MW as officially announced.
Recently, it has been shown in reference~\cite{Agarwalla:2013ju} 
that equal runs in neutrino and anti-neutrino modes
in T2K experiments are vital to settle the octant ambiguity of 
$\tz$ for all values of $\dcp$. 
Therefore, we assume equal neutrino and anti-neutrino runs of 2.5 years 
each for the T2K while exploring the octant
sensitivity. The signal efficiency in T2K is around $87\%$. In our 
simulation, the background information and other details
for T2K experiment are taken from~\cite{fechnerthesis,Huber:2009cw}.

In the NO$\nu$A experiment, the NuMI beam will be sent towards a 
14 $\mathrm{kt}$ totally active scintillator detector (TASD) placed 
at a distance of 810 $\mathrm{km}$ from Fermilab, at a location which 
is $0.8^\circ$ off-axis from the beam. Due to the off-axis location, 
the flux is sharply peaked around 2 $\mathrm{GeV}$, again close to 
the first oscillation maximum in $P(\numu \to \nue)$ channel.
The experiment is scheduled to have three years run in neutrino mode 
first and then later, three years run in anti-neutrino mode. 
The NuMI beam power is $0.7$ MW, which corresponds to $6\times 10^{20}$ 
protons on target (p.o.t.) per year. See, reference~\cite{Ayres:2007tu} 
for details. After the discovery of moderately large value of $\ty$, 
NO$\nu$A has reoptimized its event selection criteria. 
A few cuts have been relaxed to allow more events in both signal 
and background. Additional neutral current backgrounds are reconstructed 
at lower energies and can be rejected by a kinematical cut. In our simulation, we use all these new features, the details of which are given
in~\cite{Patterson:2012zs,Agarwalla:2012bv}.

\subsection{Future Generation: LBNE and LBNO}
\label{future}

The Long-Baseline Neutrino Experiment (LBNE)~\cite{Akiri:2011dv,Adams:2013qkq} 
is one of the major components of Fermilab's intensity 
frontier program. In its first phase (LBNE10), it will have a new, 
high intensity, on-axis neutrino beam directed towards a 
$10\,\mathrm{kt}$ LArTPC located at Homestake with a baseline of 
$1300\,\mathrm{km}$. This facility is designed for initial operation 
at a proton beam power of $708\,\mathrm{kW}$, with proton energy of 
$120\,\mathrm{GeV}$ that will deliver $6 \times 10^{20}$ p.o.t. in 
230 days per calendar year. In our simulation, we have used the latest 
fluxes being considered by the collaboration, which have been estimated 
assuming the smaller decay pipe and the lower horn current compared to 
the previous studies~\cite{mbishai}. We have assumed five years each 
of $\nu$ and $\bar\nu$ runs. The detector characteristics have been 
taken from Table 1 of ~\cite{Agarwalla:2011hh}. To have the LArTPC 
cross-sections, we have scaled the inclusive charged current (CC) 
cross sections of water by 1.06 (0.94) for the $\nu$ ($\bar{\nu}$) 
case~\cite{zeller,petti-zeller}.

The Long-Baseline Neutrino Oscillation Experiment (LBNO)~\cite{Stahl:2012exa} 
plans to use an experimental set-up where neutrinos produced in a conventional 
wide-band beam facility at CERN would be observed in a proposed $20\,\mathrm{kt}$
(in its first phase) LArTPC housed at the Pyh\"asalmi mine in Finland, 
at a distance of $2290\,\mathrm{km}$. The fluxes have been
computed~\cite{poster} assuming an exposure of $1.5 \times 10^{20}$ p.o.t. 
in 200 days per calendar year from the SPS accelerator at
$400\,\mathrm{GeV}$ with a beam power of $750\,\mathrm{kW}$. For LBNO also, 
we consider five years each of $\nu$ and $\bar\nu$ runs.
We assume the same detector properties as that of LBNE10. In our 
calculations, we also consider a LBNO configuration reducing the
detector mass to 10 kt which we denote as 0.5*LBNO. The exposure for this 
setup will be quite similar to LBNE10 which will enable us to
perform a comparative study between these two setups at the same footing. 
The results presented in this paper are obtained using the 
GLoBES software~\cite{Huber:2004ka,Huber:2007ji}.

\section{Physics with Bi-events Plot}
\label{bi-events-plots}

\begin{figure}[tp]
\begin{center}
\includegraphics[width=12cm, height=8.0cm]{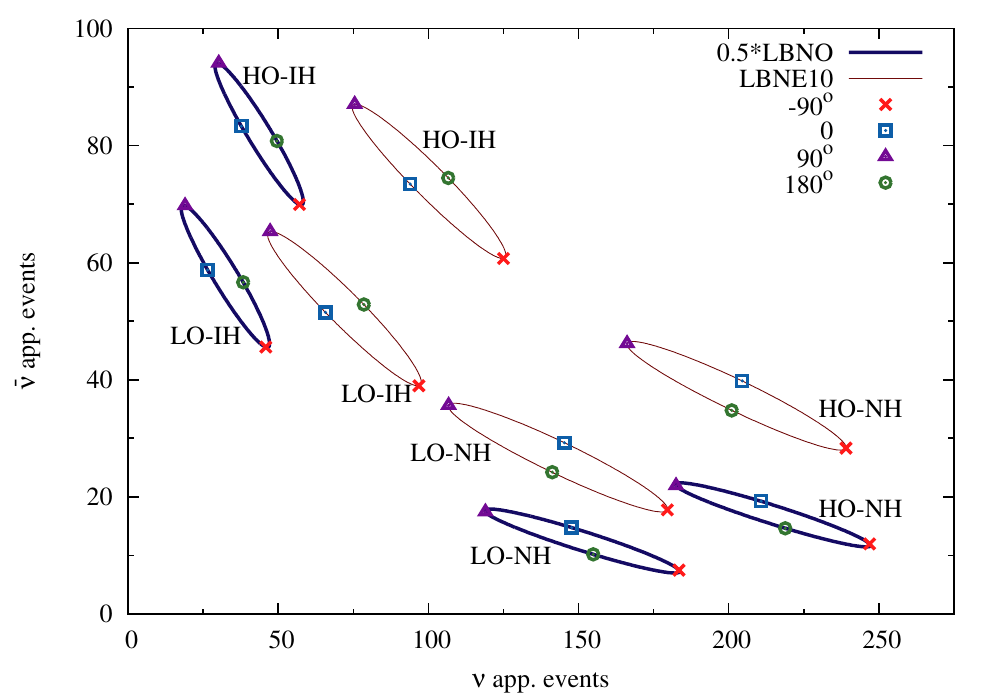}
\caption{\footnotesize{Bi-events ($\nu_e$ and $\anu_e$ appearance) plot for the four possible octant-hierarchy 
combinations and all possible $\dcp$ values. The experiments considered are LBNE10 and 0.5*LBNO.
Here $\sin^22\ty = 0.089$. For LO (HO), $\sin^2\tz = 0.41(0.59)$.}}
\label{bi-events}
\end{center}
\end{figure}

In this section, we attempt to understand the physics capabilities 
of 0.5*LBNO and LBNE10 setups with the help of bi-events plot. 
This kind of plot is quite useful to get a qualitative estimate of 
the physics sensitivity before performing a full $\Delta\chi^2$
calculation. In figure~\ref{bi-events}, we have plotted $\nu_e$ vs. 
$\bar{\nu}_e$ appearance events, for 0.5*LBNO and LBNE10
for the four possible combinations of hierarchy and octant. Since 
$\dcp$ is unknown, events are generated for the full range
$[-180^\circ, 180^\circ]$, leading to the ellipses. The event rates 
are calculated using the following oscillation parameters:
$\ms = 7.5 \times  10^{-5}$ eV$^2$, $\sin^2 \theta_{12} = 
0.3$~\cite{GonzalezGarcia:2012sz},
$\Delta{\textrm{m}}^2_{{\footnotesize \textrm{eff}}} = \pm~2.4 \times 
10^{-3}$ eV$^2$~\cite{Adamson:2013whj}, and 
$\sin^2 2 \theta_{13} = 0.089$ \cite{An:2012eh}. 
$\Delta{\textrm{m}}^2_{{\footnotesize \textrm{eff}}}$ is the 
effective mass-squared difference measured using the $\nu_\mu$
survival probability and is a linear combination of $\ma$ and $\ms$. 
The value of $\ma$ is derived from
$\Delta{\textrm{m}}^2_{{\footnotesize \textrm{eff}}}$
using the relation given in~\cite{Nunokawa:2005nx,deGouvea:2005hk}. 
This relation leads to different magnitudes of $\ma$ for NH and for IH. 
For $\sa$, we choose the two degenerate best-fit values of the
global fit~\cite{GonzalezGarcia:2012sz}: 0.41 in the lower octant 
(LO) and 0.59 in the higher octant (HO).
Note that, here we have plotted the total number of events, 
whereas the actual analysis will be done based on the
spectral information. Nevertheless, the contours in this figure 
contain very important information regarding the physics capabilities 
of the experiments. An experiment can determine both the hierarchy 
and the octant, if every point on a given ellipse is well separated 
from every point on each of the other three ellipses. The larger 
the separation, the better is the confidence level with which the 
above parameters can be determined.

One can see from figure~\ref{bi-events} that for 0.5*LBNO, 
the two (LO/HO)-IH ellipses are well separated from
the two (LO/HO)-NH ellipses, in number of $\nu_e$ events. 
Hence, 0.5*LBNO has excellent hierarchy determination 
capability with just $\nu$ data. However, $\nu$ data alone will not 
be sufficient to determine the octant in case of IH, 
because various points on (LO/HO)-IH ellipses have the same 
number of $\nu_e$ events. Likewise, only $\bar\nu$ data 
cannot determine the octant in case of NH. Therefore, balanced 
$\nu$ and $\bar\nu$ data are mandatory to make an 
effective distinction between (LO/HO)-IH ellipses and also 
between (LO/HO)-NH ellipses. Figure~\ref{bi-events} also
depicts that the asymmetries between the neutrino and 
anti-neutrino appearance events are largest for the combinations:
(NH, $\dcp = -90^\circ$) and (IH, $\dcp = 90^\circ$). 

For LBNE10, $\nu$ data alone can not determine hierarchy because 
various points on LO-NH and HO-IH ellipses
have the same number of $\nu_e$ events (see figure~\ref{bi-events}). 
Thus, $\bar\nu$ data is also needed.
Even with $\bar\nu$ data, hierarchy determination can be difficult 
to achieve, if nature chooses LO and one of the two worst 
case combinations of hierarchy and $\dcp$ which are (NH, $90^\circ$) 
or (IH, $-90^\circ$).
In such a situation, the $\nu_e$ and $\bar\nu_e$ events are rather 
close to each other and it will be very difficult for LBNE10 
to reject the wrong combination. Regarding octant determination, 
the capability of LBNE10 is very similar to that of 0.5*LBNO
because the separations between the ellipses, belonging to LO and 
HO are very similar for these two experiments.

\section{Our Findings}
\label{results}

Measurement of mass hierarchy and octant should be considered 
as a prerequisite for the discovery of leptonic
CP violation. Now, it would be quite interesting to study whether 
the expected appearance data from the first phases of
LBNE and LBNO experiments can resolve the issues of neutrino mass 
hierarchy and octant of $\tz$ at 
3$\sigma$ to 5$\sigma$ confidence level before they start probing 
the parameter space for leptonic CP violation.
In this section, we address these issues in detail. We present 
the results for LBNE10 (10 kt), 0.5*LBNO (10 kt), 
and LBNO (20 kt) setups. We also study the improvement in
their physics reach when the projected data from current 
generation experiments T2K and NO$\nu$A, is added. 
The impact of T2K and NO$\nu$A measurements on the 
performance of LBNE setup to determine the mass hierarchy
and discover leptonic CP violation 
has been discussed recently in~\cite{Blennow:2013swa}.

\subsection{Discovery Reach for Neutrino Mass Hierarchy}
\label{results-mh}

We first focus on the discovery potential of future facilities 
to exclude the wrong hierarchy. It can be seen from 
equation~\ref{eq:matt} that the first term ($C_0$) dominates 
for large $\ty$ and it is the leading term in platinum channel.
This term contains the largest Earth matter effect which can 
therefore be used to unravel the sign of $\ma$. This term is
also proportional to $\sa$ and therefore is quite sensitive to 
the choice of $\tz$ value. If we vary $\sa$ in its 3$\sigma$
allowed range of 0.34 to 0.67, then for LBNE10, the signal 
event rates in $\nue$ appearance channel will increase 
from 122 to 231 (assuming NH and $\dcp = 0^\circ$), an almost 
$\sim$ 90\% enhancement in the statistics. 
For LBNO setup with 20 kt detector size, these numbers will 
change from 247 to 478 showing an almost $\sim$ 94\%
increase in the event numbers. $\Delta\chi^2$ is calculated 
for a given true combination of $\tz$-hierarchy,
assuming the opposite hierarchy to be the test hierarchy. 
In the fit, we marginalize over test $\sin^2\theta_{23}$
in its $3\sigma$ range. 
$\Delta{\textrm{m}}^2_{{\footnotesize \textrm{eff}}}$ and 
$\sin^22\theta_{13}$ are marginalized in their $2\sigma$
ranges. We consider 5$\%$ uncertainty in the matter density, 
$\rho$. Priors were added for $\rho$ ($\sigma=5\%$), 
$\Delta{\textrm{m}}^2_{{\footnotesize \textrm{eff}}}$ 
($\sigma=4\%$), and $\sin^22\ty$ ($\sigma=5\%$, as expected by 
the end of Daya Bay's run~\cite{dayabay_NF12}). 
$\dchsq$ is also marginalized over the uncorrelated 
systematic uncertainties (5\% on signal and 5\% on background) 
in the set-ups, so as to obtain a 
$\dchsq_{{\footnotesize \textrm{min}}}$ for every $\dcp$(true).

\begin{figure}[tp]
\begin{center}
\includegraphics[width=12cm, height=8.0cm]{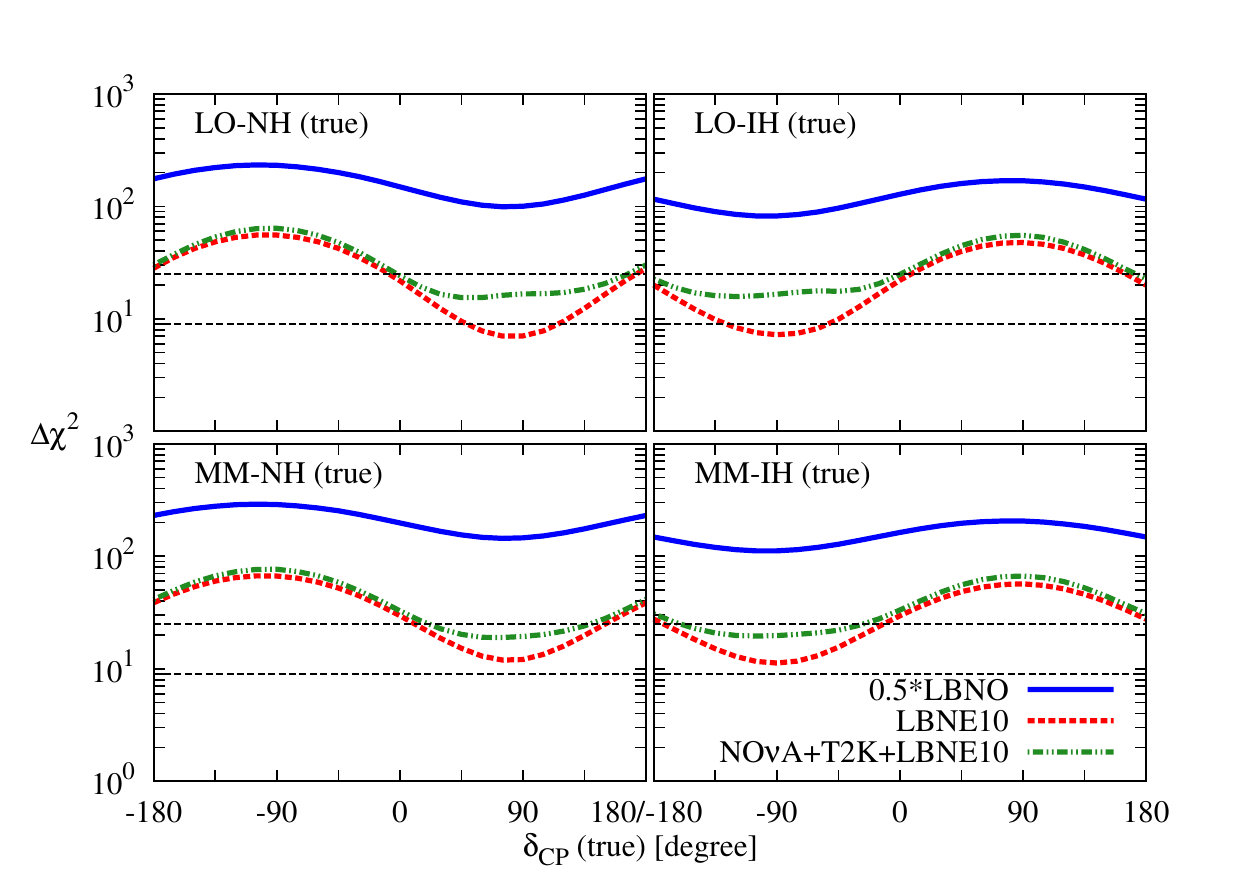}
\caption{\footnotesize{Discovery reach for mass hierarchy as a function of true $\dcp$ for 0.5*LBNO, LBNE10, and 
LBNE10 combining the projected data from T2K and NO$\nu$A (see section~\ref{experiments}). Results are shown 
for four possible true $\tz$-hierarchy combinations. For LO (MM), $\sin^2\tz(\textrm{true}) = 0.41~(0.5)$. Here 
$\sin^22\ty(\textrm{true}) = 0.089$.}}
\label{hierarchy_exclusion_bestfit}
\end{center}
\end{figure}

First, we consider two true values of $\sin^2\tz$: 
0.41 (best-fit value in LO) and 0.5 (MM) giving us four 
true combinations of $\tz$-hierarchy: LO-NH, LO-IH, MM-NH 
and MM-IH. The hierarchy reach would suffer the most if 
$\sa$(true) belongs to LO,  hence we show the results 
for the best-fit value in LO. Here, we would like to 
mention that if we take $\sa$(true) to be the 
best-fit value in HO, then the discovery reaches of these 
experiments will be better than that for the case of MM.
We elaborate on this point at the end of this section. 
Figure~\ref{hierarchy_exclusion_bestfit} depicts the 
discovery reach for hierarchy as a function 
of $\dcp$(true). We see that even 0.5*LBNO has 
$\gtrsim 10\sigma$\footnote{To estimate this, we use the relation 
$\textrm{n}\sigma = \sqrt{\dchsq_{{\footnotesize \textrm{min}}}}$. 
However, in order to calculate the sensitivity to the mass hierarchy, 
a new method has been described in ~\cite{Qian:2012zn} considering 
the fact that a discrete parameter does not follow a 
Gaussian distribution.} hierarchy discovery potential for 
all values of $\dcp$(true) and for all four true $\tz$-hierarchy 
combinations. The potential of LBNO, of course, is even better. 
The LBNO baseline is close to bimagic which gives it a particular
advantage~\cite{Raut:2009jj,Dighe:2010js}. For LBNE10, a 5$\sigma$ 
discovery of hierarchy is possible for only $\sim50\%$ 
of the $\dcp$(true), irrespective of these four true 
$\tz$-hierarchy combinations. For the unfavorable hierarchy-$\dcp$ 
combinations~\cite{Prakash:2012az}, {\it i.e.} NH with $\dcp$ in the upper half plane or IH with 
$\dcp$ in the lower half plane, the performance of LBNE10 suffers. 
In particular, for LO and the worst case combinations 
[(NH, $90^\circ$) and (IH, $-90^\circ$)], LBNE10 will not be
able to provide even a 3$\sigma$ hierarchy discrimination. 
This suggests that additional data is needed for LBNE10 to 
have such a capability. In such a scenario, the projected data 
from T2K and NO$\nu$A can come to the rescue. Adding data from 
T2K (5 years of neutrino run) and NO$\nu$A (3 years of $\nu$ run 
and 3 years of $\anu$ run) helps LBNE10 setup to achieve more 
than 3$\sigma$ discovery reach for mass hierarchy 
irrespective of the true choices of hierarchy
and $\dcp$ (see upper panels of figure~\ref{hierarchy_exclusion_bestfit}),
even if $\tz$ is in the lower octant. 

\begin{figure}[tp]
\centering
\includegraphics[width=0.49\textwidth]{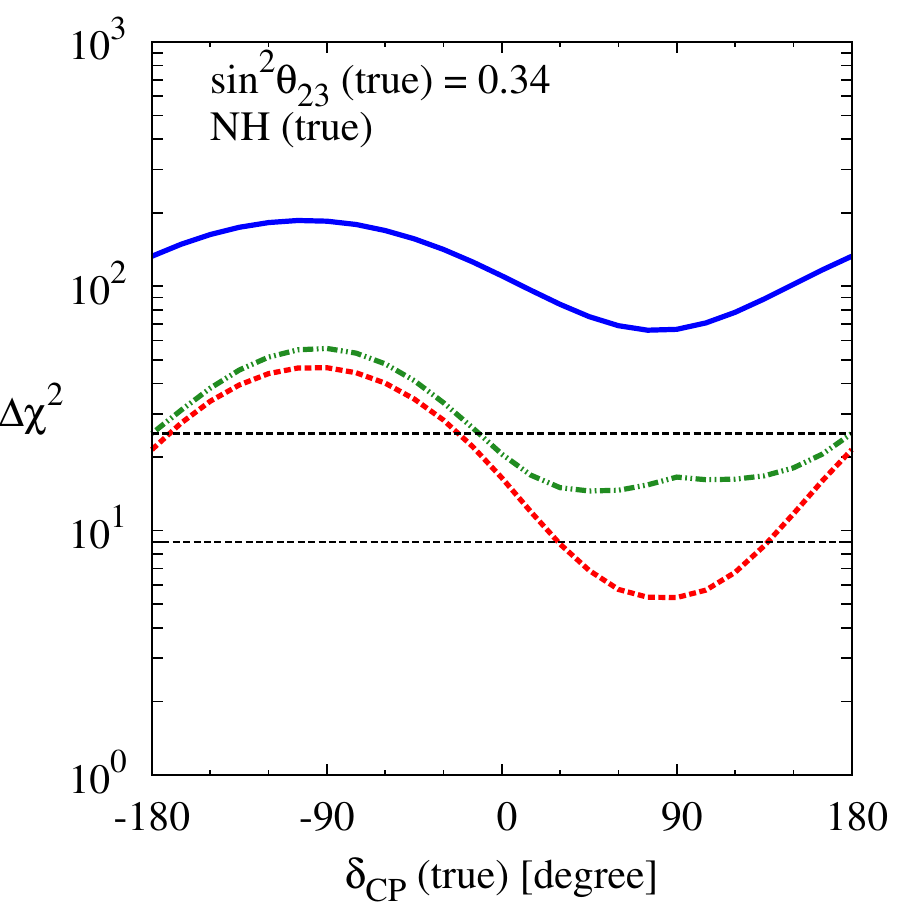}
\includegraphics[width=0.49\textwidth]{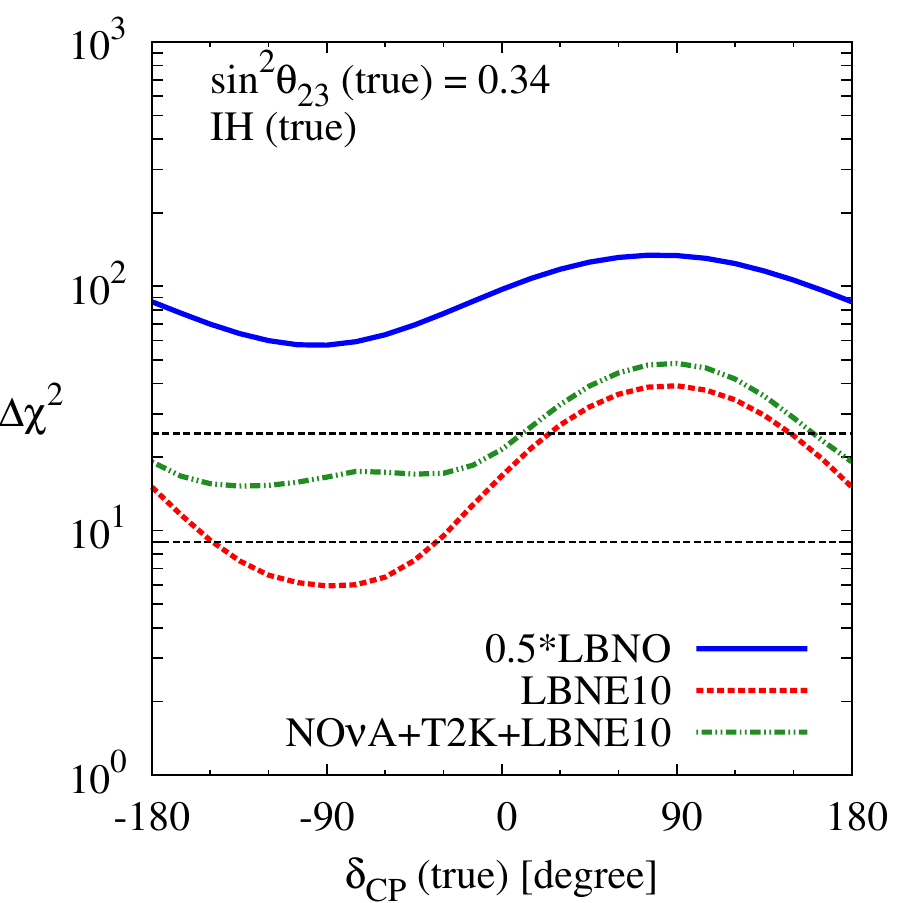}
\caption{\footnotesize{Left panel (right panel) shows the $\Delta\chi^2$ for the mass hierarchy discovery as a function of true
value of $\dcp$ assuming NH (IH) as true hierarchy. Results are shown for 0.5*LBNO, LBNE10, and LBNE10+T2K+NO$\nu$A.
Here we consider $\sa$(true) = 0.34 (the lowest value in its allowed 3$\sigma$ range).}}
\label{hierarchy_exclusion_ssqt23_lowest_3sigma}
\end{figure}

Now, we ask the question, by how much does the sensitivity 
deteriorate if $\sa$(true) turns out to be 0.34 in nature,
 which is its minimum value allowed in the 3$\sigma$ range? 
We have checked that even in this case, LBNO setup with 20 kt 
detector mass can give $\dchsq_{{\footnotesize \textrm{min}}} 
\gtrsim 100$ irrespective of the true choices of hierarchy,
and $\dcp$.
From figure~\ref{hierarchy_exclusion_ssqt23_lowest_3sigma}, 
it can be seen that 0.5*LBNO can resolve the issue of mass
hierarchy at more than 7$\sigma$ confidence level for 
$\sa$(true) = 0.34 independent of the choices of true hierarchy 
and $\dcp$. The most important message that is conveyed by 
figure~\ref{hierarchy_exclusion_ssqt23_lowest_3sigma} is that 
with the help of projected T2K and NO$\nu$A data, LBNE10 can 
still achieve 3$\sigma$ mass hierarchy discovery for any 
combinations of true hierarchy-$\dcp$-$\sa$.
It clearly demonstrates the synergy between the current 
(off-axis) and future (on-axis) superbeam experiments and 
also proves that adding data from three different baselines 
(295 km, 810 km, and 1300 km) with completely different energy 
spectra is quite useful to kill the clone solutions for the 
unfavorable choices of the oscillation parameters. 

The mass hierarchy discovery potential for all the three set-ups 
under consideration is remarkable if $\tz$ happens to lie in HO. 
For $\sa$(true) = 0.59 (the best-fit value in HO), even 0.5*LBNO 
can have $\dchsq_{{\footnotesize \textrm{min}}} \gtrsim 130$ 
irrespective of the true choices of hierarchy and $\dcp$. With this 
choice of $\sa$(true), a 5$\sigma$ discovery is not possible with LBNE10
for $\sim30\%$ values of true $\dcp$ in the upper half plane for NH 
true and for $\sim70\%$ values of true $\dcp$ in the lower 
half plane for IH true. We have checked that if we add the data 
from T2K and NO$\nu$A, LBNE10 can again provide 5$\sigma$ 
discovery for mass hierarchy irrespective of the choices of true 
hierarchy and $\dcp$ with $\sa$(true) = 0.59. 
Next we turn our attention to the octant discovery potential 
of these setups.

\subsection{Discovery Reach for $\tz$ Octant}
\label{results-octant}

Here we discuss the discovery reach of future facilities 
for excluding the wrong octant. We consider the best-fit 
true values of $\sin^2\tz=0.41$ (in LO) and 0.59 (in HO) 
resulting into the following four true combinations of 
octant and hierarchy: LO-NH, LO-IH, HO-NH, and HO-IH. 
$\Delta\chi^2$ is calculated for each of these four combinations, 
assuming test $\sin^2\tz$ values from the other octant. For 
LO (HO) true, we consider the test $\sin^2\tz$ range from 
0.5 to 0.67 (0.34 to 0.5). Rest of the marginalization 
procedure (over other oscillation parameters and systematic 
uncertainties) is the same as that in the case of hierarchy 
exclusion except with another difference: the final $\dchsq$ 
is marginalized over both the hierarchies in the fit to obtain 
$\dchsq_{{\footnotesize \textrm{min}}}$.

\begin{figure}[tp]
\begin{center}
\includegraphics[width=12cm, height=8.0cm]{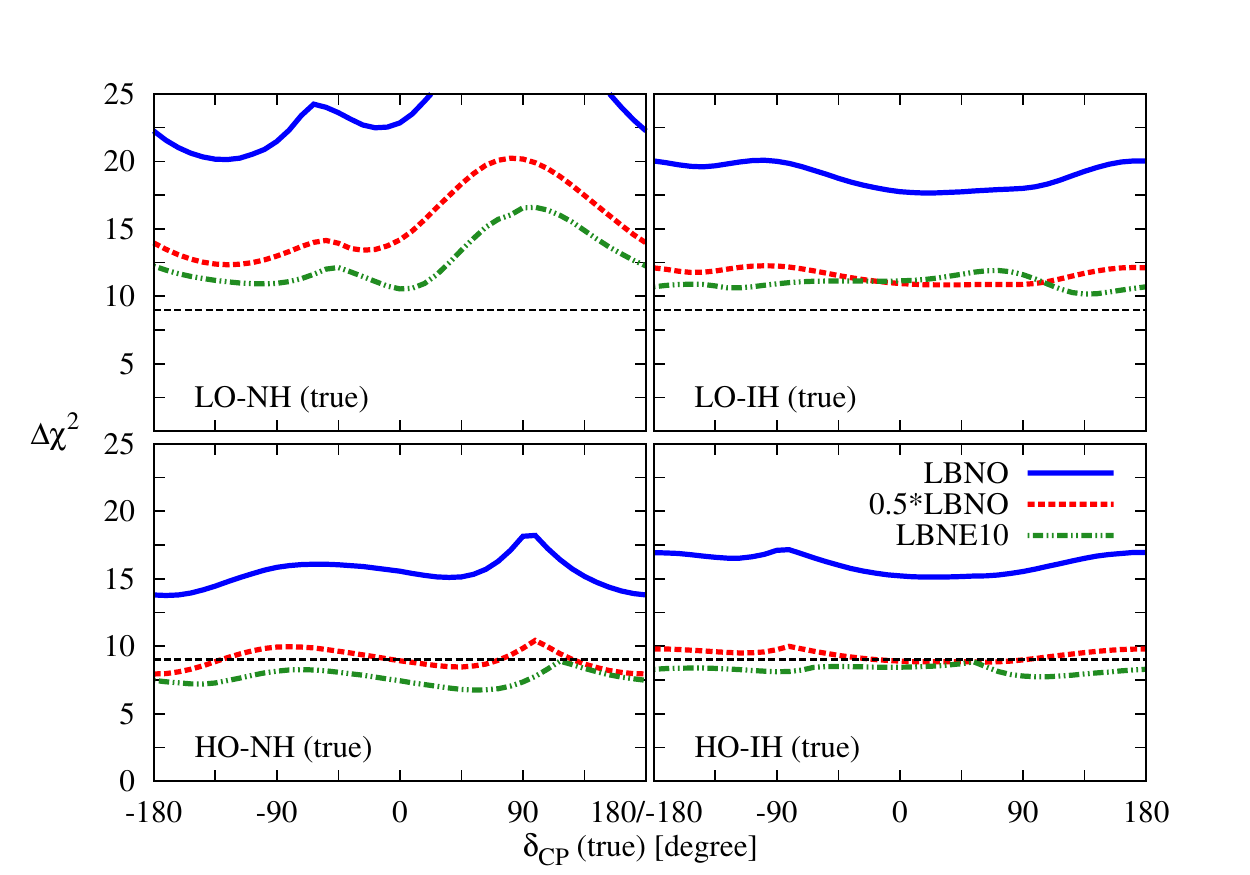}
\caption{\footnotesize{Octant resolving capability as a function of true $\dcp$
for LBNO, 0.5*LBNO, and LBNE10. Results are shown for the four possible true octant-hierarchy
combinations. For LO (HO), $\sin^2\tz(\textrm{true}) = 0.41~(0.59)$. Here $\sin^22\ty(\textrm{true}) = 0.089$. }}
\label{octant_exclusion_futurefacilities}
\end{center}
\end{figure}

Figure~\ref{octant_exclusion_futurefacilities} shows the 
discovery reach for octant as a function of $\dcp$(true).
It can be seen that for (LO/HO)-IH true, the sensitivities 
of LBNE10 and 0.5*LBNO are quite similar whereas they are
somewhat better for 0.5*LBNO if (LO/HO)-NH are the true 
combinations. For LO-(NH/IH), both LBNE10 and 0.5*LBNO
have more than $3\sigma$ discovery of octant while for 
HO-(NH/IH), the $\dchsq_{{\footnotesize \textrm{min}}}$ 
varies from $\sim$ 7 to 11 depending on the true value of 
$\dcp$. However, with full LBNO, we have more than $3.5\sigma$ 
discovery of octant for all true octant-hierarchy-$\dcp$ 
combinations. A $5\sigma$ discovery of octant is possible 
only for LO-NH true for $\dcp(\textrm{true})$ $\in$ 
($\sim 20^\circ$ to $150^\circ$).

\begin{figure}[tp]
\begin{center}
\includegraphics[width=12cm, height=8.0cm]{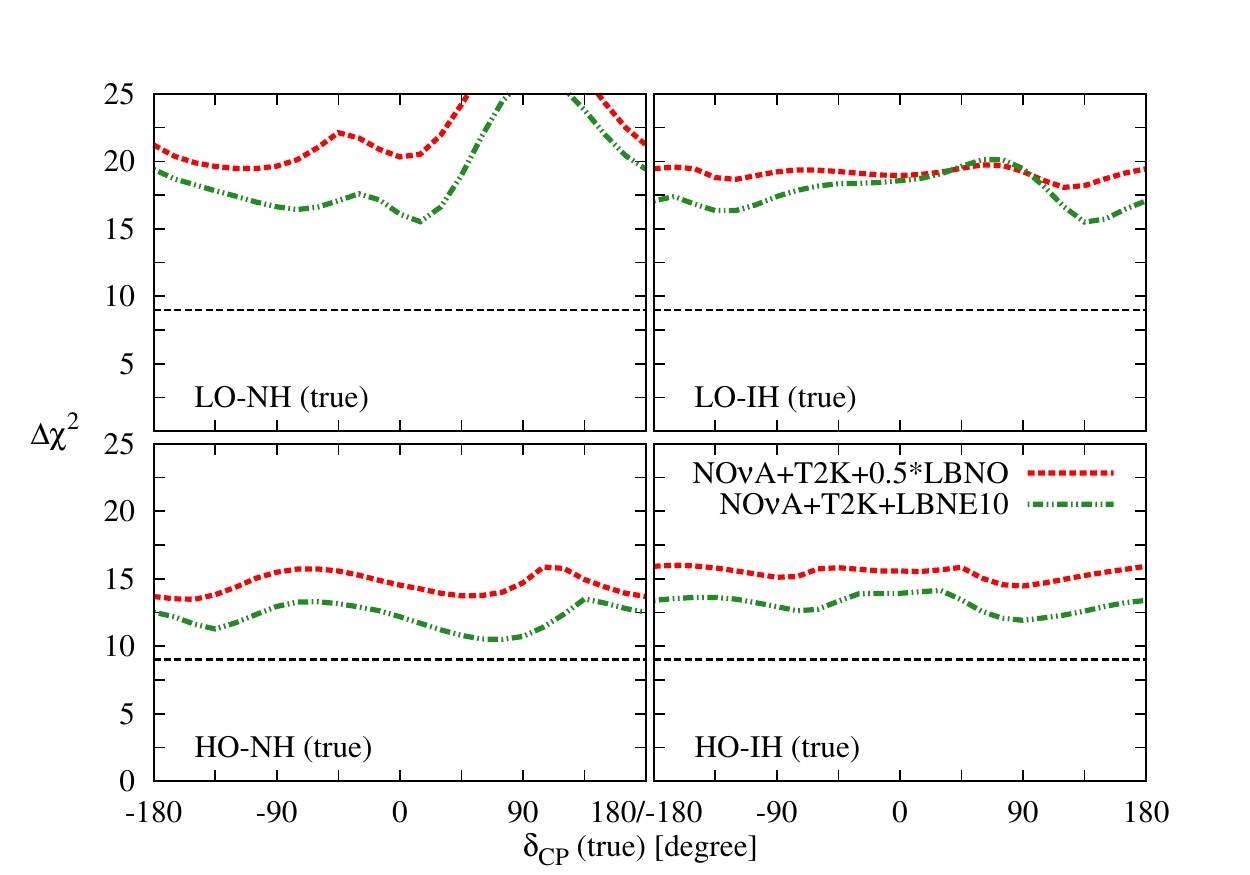}
\caption{\footnotesize{$\dchsq_{{\footnotesize \textrm{min}}}$ for octant discovery potential as a function of true $\dcp$
for 0.5*LBNO and LBNE10 adding the projected data from T2K and NO$\nu$A. Results are shown for the four possible true 
octant-hierarchy combinations. For LO (HO), $\sin^2\tz(\textrm{true}) = 0.41~(0.59)$. Here $\sin^22\ty(\textrm{true}) = 0.089$.}}
\label{octant_exclusion_presentplusfuturefacilities}
\end{center}
\end{figure}

In figure~\ref{octant_exclusion_presentplusfuturefacilities}, 
we present the improvement in the octant discovery reach 
for 0.5*LBNO and LBNE10 with the addition of the projected 
data from T2K (2.5 years of $\nu$ run and 2.5 years of $\anu$ 
run) and NO$\nu$A (3 years of $\nu$ run and 3 years of $\anu$ 
run). Adding data from current generation experiments helps 
both 0.5*LBNO and LBNE10 to achieve more than 3$\sigma$ 
discovery for all true octant-hierarchy-$\dcp$ combinations. 
For LO-(NH/IH) true, these setups can provide close to 
3.8$\sigma$ discovery for octant irrespective of the choice of true $\dcp$.

\begin{figure}[tp]
\centering
\includegraphics[width=0.49\textwidth]{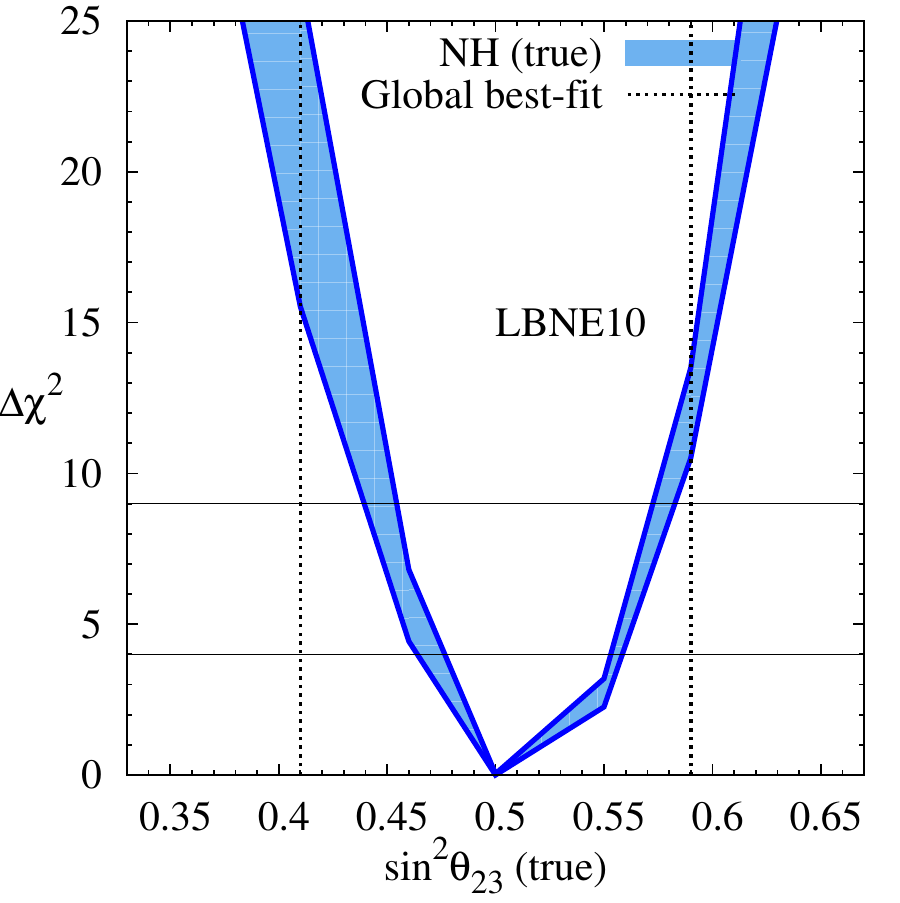}
\includegraphics[width=0.49\textwidth]{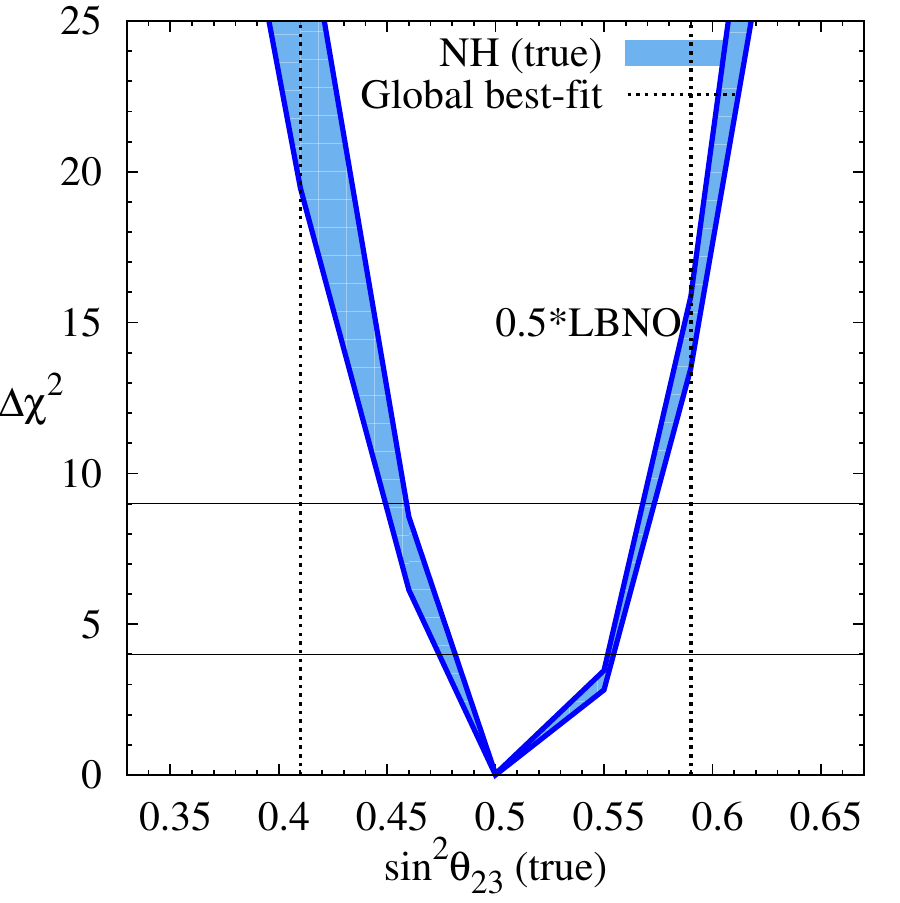}
\caption{\footnotesize{$\dchsq_{{\footnotesize \textrm{min}}}$ for octant resolution as a function of true $\sa$. Left panel (right panel) is for 
LBNE10 (0.5*LBNO). The variation due to $\dcp$(true) leads to the band in $\dchsq$ for a given $\sa$(true).
The vertical lines correspond to the global best-fit values. We consider NH as true hierarchy. In producing all these plots, 
the projected data from T2K and NO$\nu$A have been added (see section~\ref{experiments} for details).}}
\label{dchsq_vs_true_tz_nh_true}
\end{figure}

In the discussion so far, we consider only the best-fit 
true values of $\sa$ in both the octants. Now, we address the 
octant resolution capability for values of true $\sa$ in the 
full 3$\sigma$ allowed range of 0.34 to 0.67. In 
figure~\ref{dchsq_vs_true_tz_nh_true}, we plot the 
$\dchsq_{{\footnotesize \textrm{min}}}$ as a function of true 
$\sa$ for LBNE10 (left panel) and 0.5*LBNO (right panel) 
assuming NH as true hierarchy. Variation of $\dcp$(true) in 
the range $-180^\circ$ to $180^\circ$ leads to the band in 
$\dchsq$ values for a given true $\sa$. The vertical lines 
point towards the global best-fit values. Here we have added 
the projected data from T2K and NO$\nu$A to produce these 
results. For LBNE10, a 3$\sigma$ octant resolution is possible 
for $\sa$(true) $\leq$ 0.44 and for $\sa$(true) $\geq$ 0.58 for 
all values of $\dcp$(true). For 0.5*LBNO, this is possible for 
$\sa$(true) $\leq$ 0.45 and for $\sa$(true) $\geq$ 0.57. We 
present the results for IH as true choice in appendix~\ref{appendix1}. 

\begin{figure}[tp]
\centering
\includegraphics[width=0.49\textwidth]{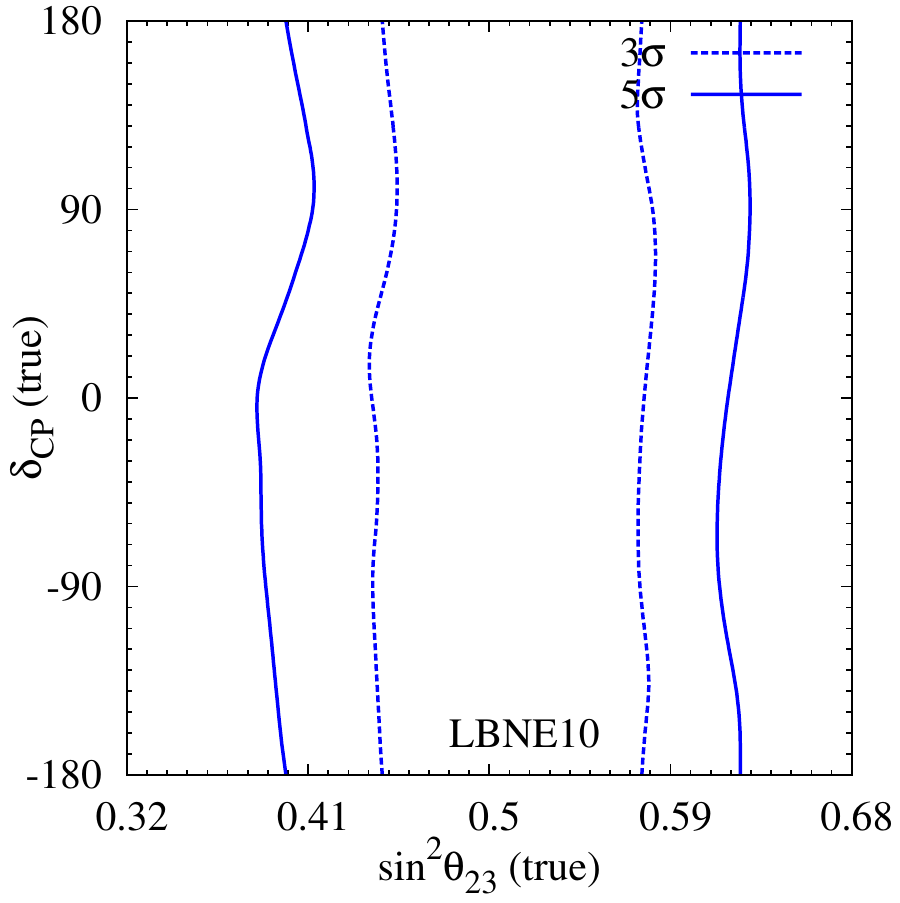}
\includegraphics[width=0.49\textwidth]{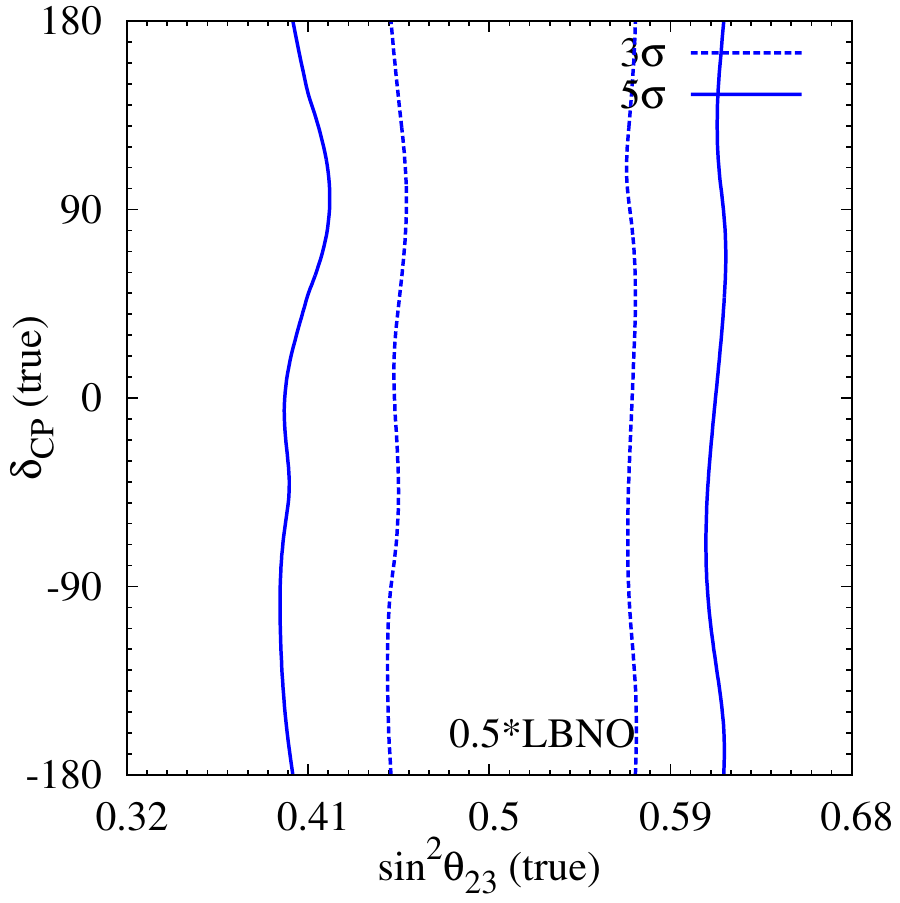}
\caption{\footnotesize{Octant resolving capability at 3$\sigma$ and 5$\sigma$ C.L. in the true $\sa$ - true $\dcp$ plane for 
LBNE10 (left panel) and 0.5*LBNO (right panel). The vertical lines point towards the global best-fit values. 
Here, we assume NH as true hierarchy. In generating all these plots, the projected data from 
T2K and NO$\nu$A have been added (see section~\ref{experiments} for details).}}
\label{true-tz-true-cp-nh-true}
\end{figure}

Figure~\ref{true-tz-true-cp-nh-true} depicts the 3$\sigma$ and 
5$\sigma$ octant resolution contours in true $\sa$ - true $\dcp$ 
plane assuming NH as true hierarchy. The left (right) panel is 
for LBNE10 (0.5*LBNO) adding the expected data from T2K and 
NO$\nu$A. Octant resolution is only possible for points lying 
outside the contours. This figure again confirms that both 
LBNE10 and 0.5*LBNO in combination with T2K and NO$\nu$A data 
can provide octant discovery for global best-fit points at 
3$\sigma$ confidence level. We show the similar figure for the 
true IH choice in appendix~\ref{appendix1}.

\subsection{Discovery Reach for Leptonic CP Violation}
\label{results-octant}

A `discovery' of leptonic CP violation, if it exists in Nature, 
means that we can reject both the CP-conserving values of 
$0^{\circ},\,180^{\circ}$ at a given confidence level. Obviously, 
this measurement becomes very difficult when $\dcp$ approaches 
to $0^{\circ},\,180^{\circ}$. Therefore, whilst it is possible 
to discover the mass hierarchy for \emph{all} possible values of 
$\dcp$, the same is not true in the case of CP violation study. 
We have already emphasized that the present uncertainty in the knowledge 
of $\sa$ has a crucial impact on the discovery reach of mass ordering 
and octant of $\tz$ for the experimental setups under consideration. 
This is also true for the CP violation discovery reach. We can see 
from the appearance probability expression in equation~\ref{eq:matt} that both the 
CP-violating ($C_-$) and CP-conserving ($C_+$) terms depend on 
$\sin2\theta_{23}$, therefore these terms are not sensitive to the
octant of $\tz$ but they depend on the value of $\tz$. The leading 
term ($C_0$) in equation~\ref{eq:matt} is proportional to $\sa$ and 
therefore it is sensitive to both the octant and magnitude of $\tz$. 
In this paper for the first time, we study in detail the CP violation 
discovery reach by varying the true value of $\sa$ in its allowed 
3$\sigma$ range of 0.34 to 0.67. We follow the same marginalization 
scheme in the fit for oscillation parameters and systematic 
uncertainties as that in the case of mass hierarchy discovery study. 
For CP violation searches, the final $\dchsq$ is also marginalized 
over both the choices of hierarchy in the fit to obtain 
$\dchsq_{{\footnotesize \textrm{min}}}$.

\begin{figure}[tp]
\centering
\includegraphics[width=0.325\textwidth]{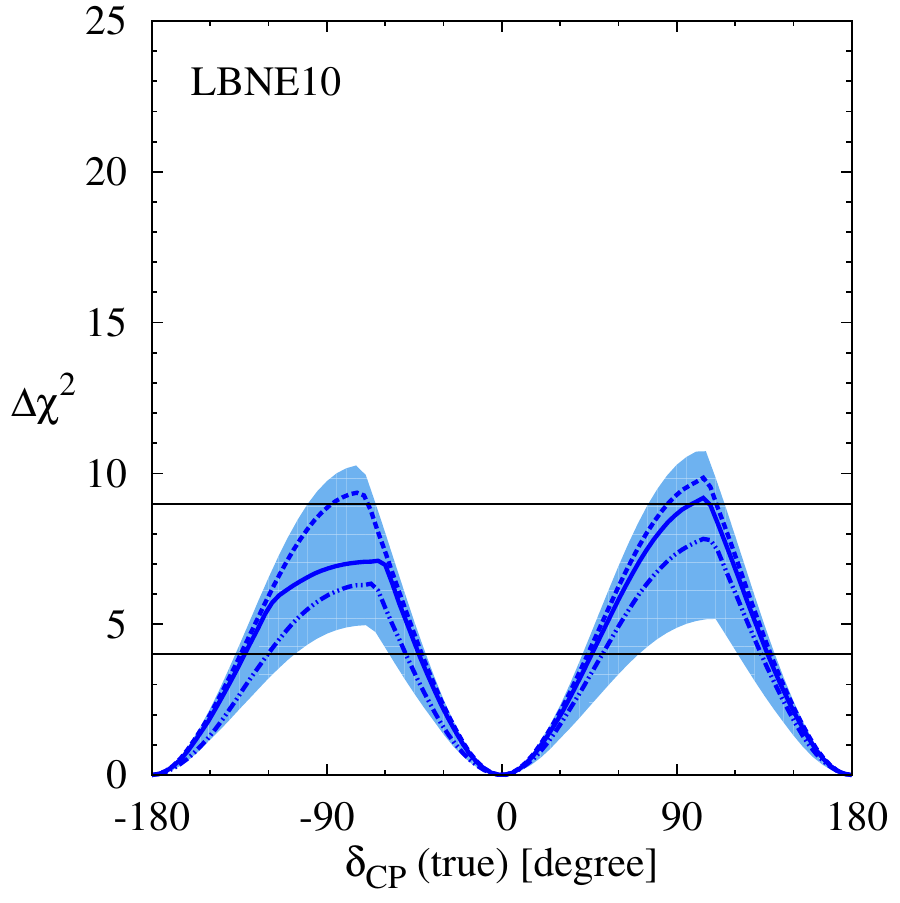}
\includegraphics[width=0.325\textwidth]{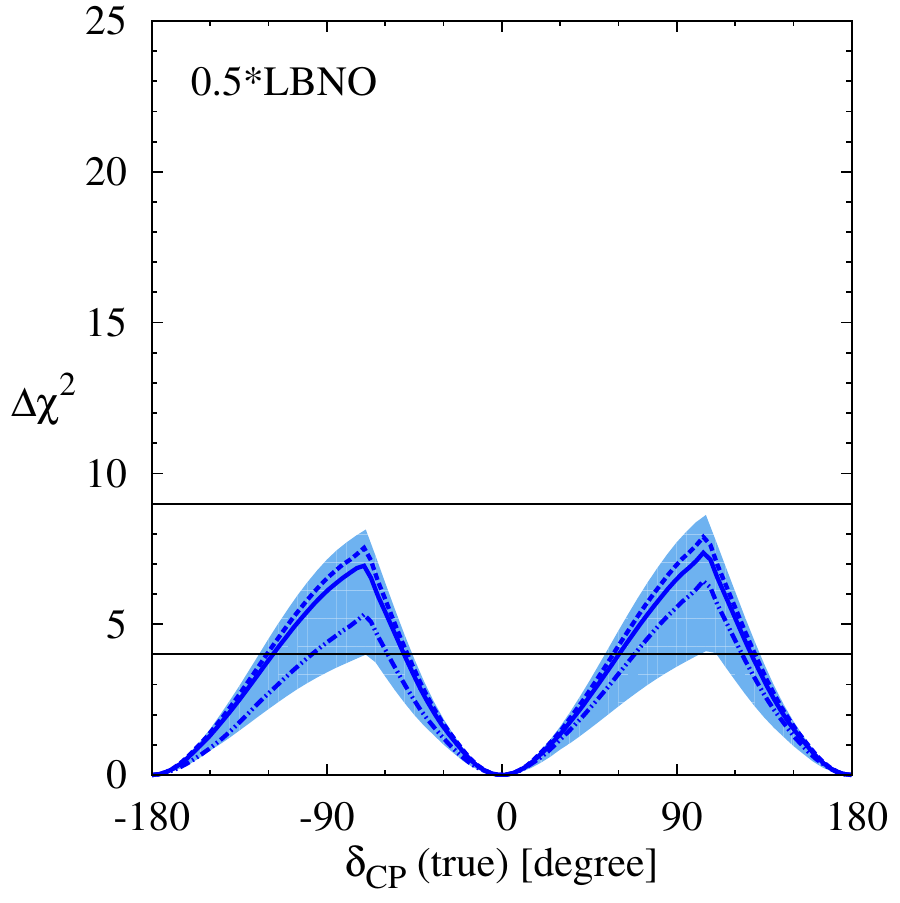}
\includegraphics[width=0.325\textwidth]{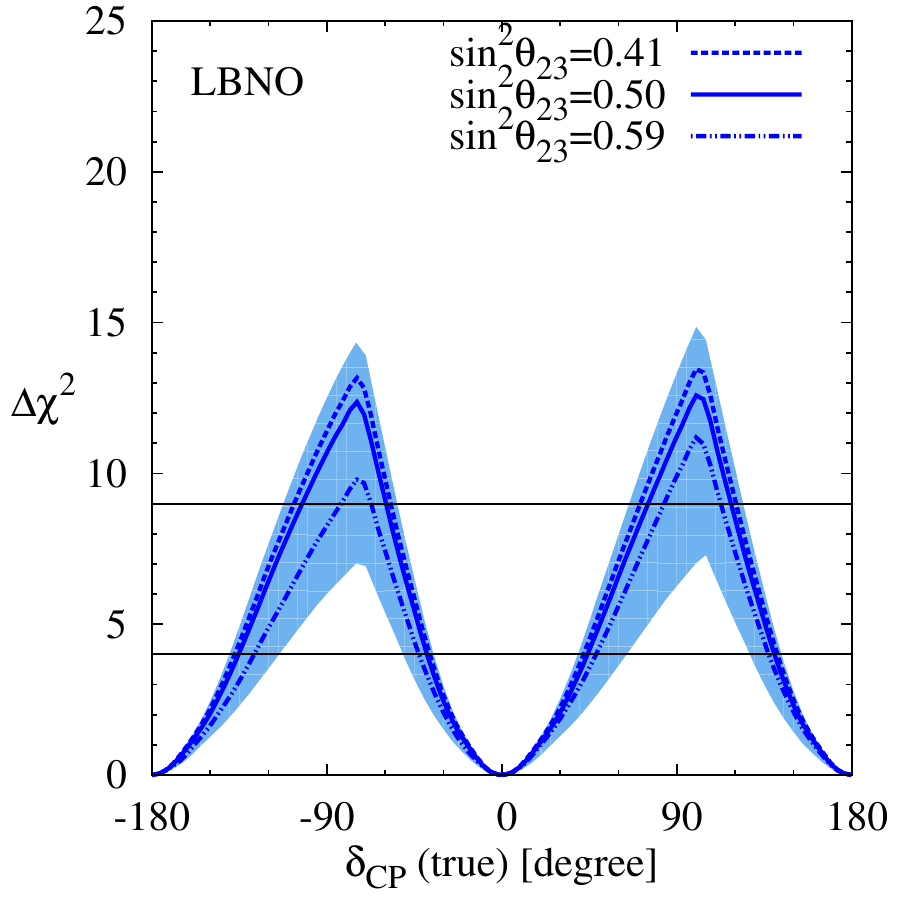}
\vskip0.5cm
\includegraphics[width=0.325\textwidth]{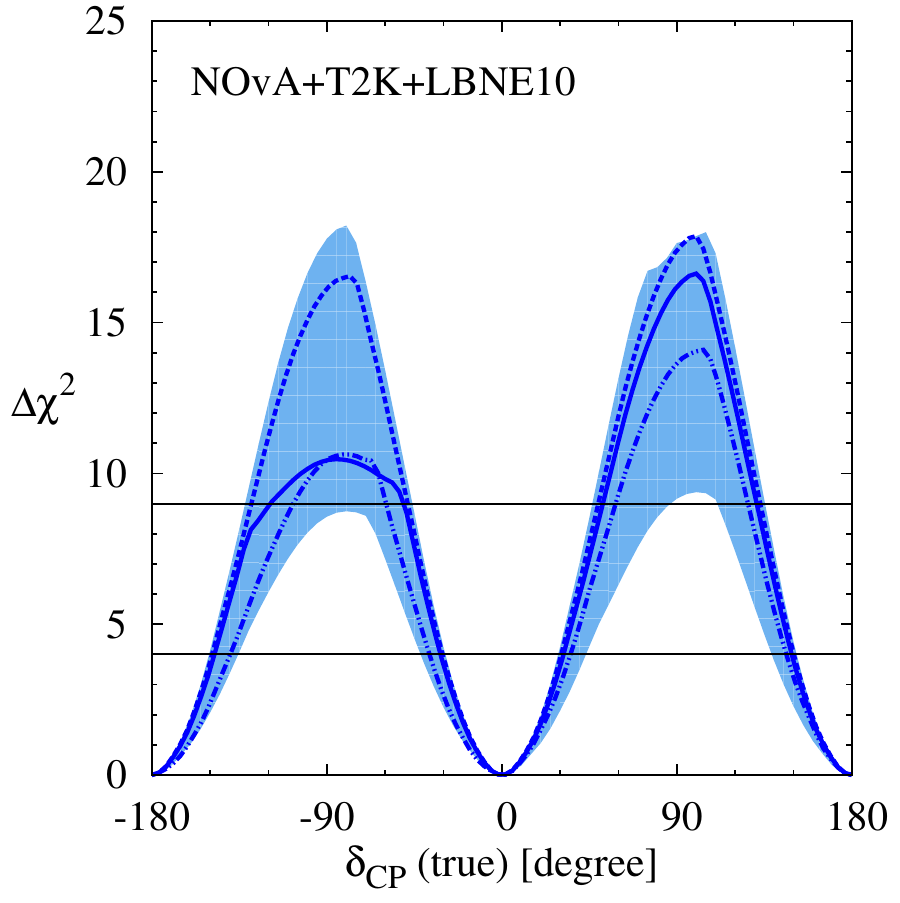}
\includegraphics[width=0.325\textwidth]{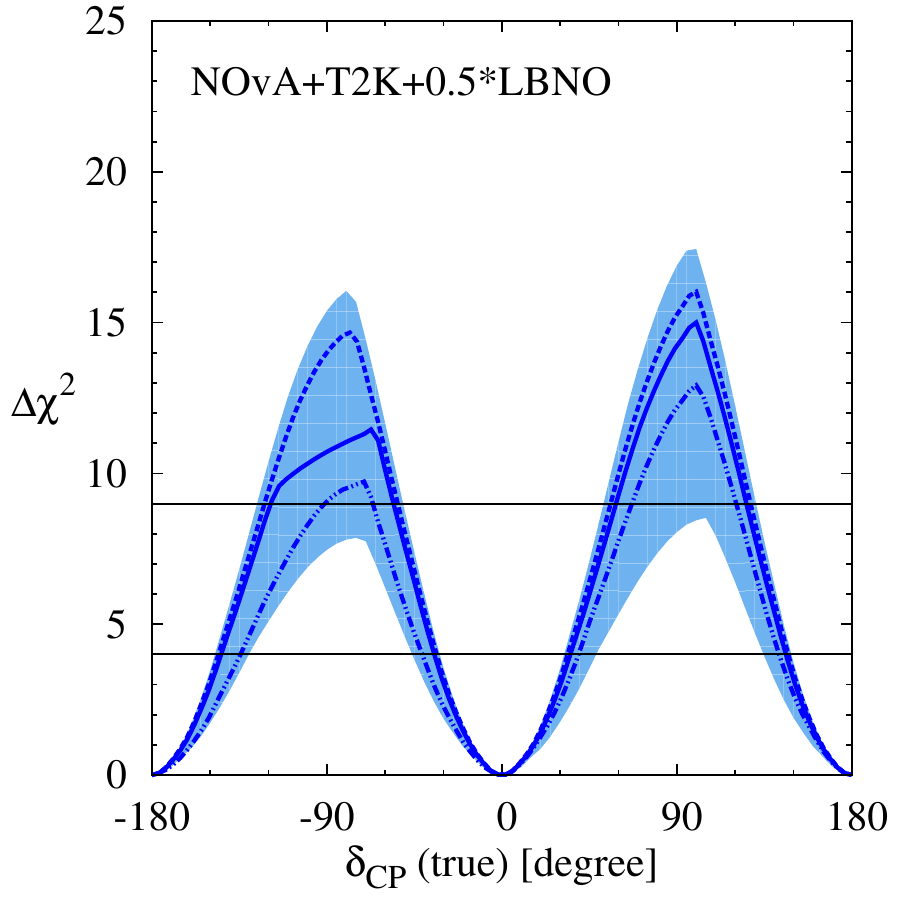}
\includegraphics[width=0.325\textwidth]{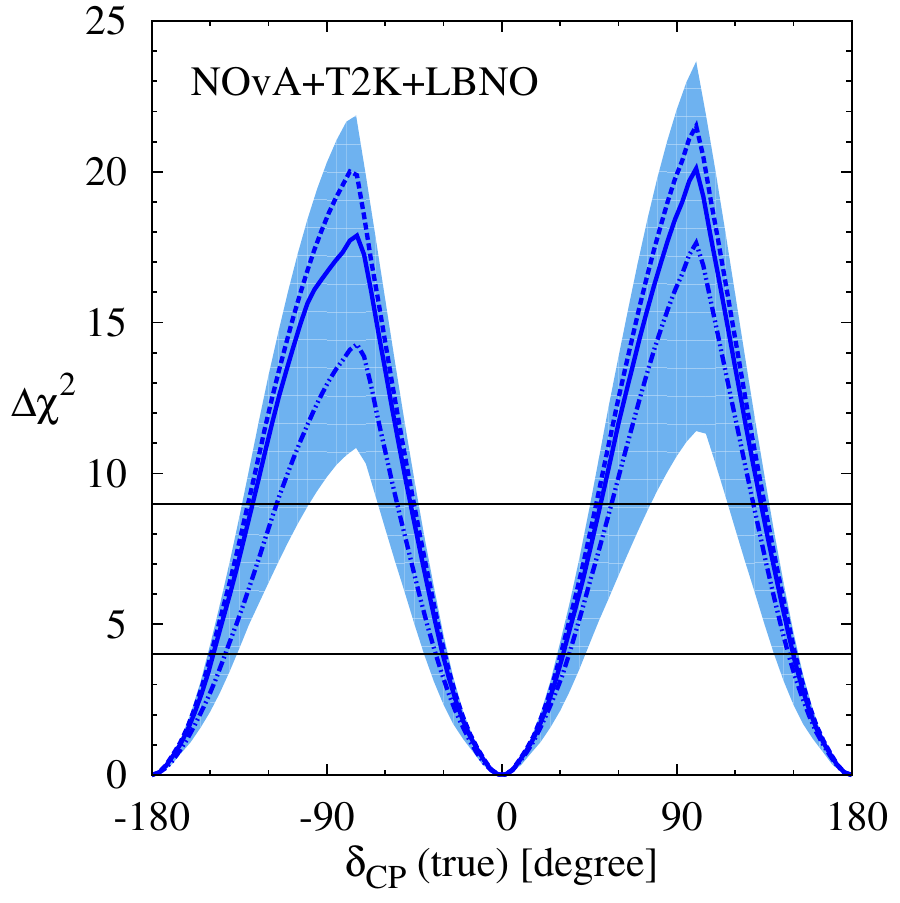}
\caption{\footnotesize{CP Violation discovery reach as a function of true value of $\dcp$ assuming NH as true hierarchy. 
Results are shown for LBNE10 (10 kt), 0.5*LBNO (10 kt), and LBNO (20 kt) setups in the left, middle, and right upper panels respectively. 
In lower panels, we show the same including the projected data from T2K and NO$\nu$A experiments. The shaded band depicts
the variation in $\dchsq_{{\footnotesize \textrm{min}}}$ due to different true choices of $\sa$ in its 3$\sigma$ allowed range of 
0.34 to 0.67. Inside the band, we show the results for three different true values of $\sa$: 0.41, 0.5, and 0.59.}}
\label{cpv-nh-lbne10-lbno10-lbno20-nova-t2k}
\end{figure}

In figure~\ref{cpv-nh-lbne10-lbno10-lbno20-nova-t2k}, we present 
the CP violation discovery reach for various experimental setups 
under consideration as a function of true $\dcp$ assuming NH as 
true hierarchy. Similar figure for the true IH choice is given 
in appendix~\ref{appendix2}. The left, middle, and right upper 
panels of figure~\ref{cpv-nh-lbne10-lbno10-lbno20-nova-t2k} show 
the results for LBNE10, 0.5*LBNO, and LBNO respectively. In lower 
panels, we depict the same results, combining the projected data from T2K 
and NO$\nu$A experiments. The shaded band in each panel reflects 
the variation in $\dchsq_{{\footnotesize \textrm{min}}}$ due to 
different true choices of $\sa$ in its 3$\sigma$ allowed range of 
0.34 to 0.67. Inside the band, we give the results for three different 
true values of $\sa$: 0.41, 0.5, and 0.59. We summarize the main 
features of figure~\ref{cpv-nh-lbne10-lbno10-lbno20-nova-t2k} in 
Table~\ref{tab:CPV-compare-LBNE10-LBNO}. In their first phases,
 both LBNE10 and LBNO will have CP violation reach for around 
50\% values of true $\dcp$ at $2\sigma$ confidence level 
(see Table~\ref{tab:CPV-compare-LBNE10-LBNO}). At $3\sigma$, 
their CP violation reach is quite minimal: only 3\% for LBNE10 
and 23\% for LBNO. It is quite important to note that the addition 
of the projected T2K and NO$\nu$A data helps a lot to improve the
CP coverage for these setups at $3\sigma$ confidence level for all 
possible true values of $\sa$ (see 
figure~\ref{cpv-nh-lbne10-lbno10-lbno20-nova-t2k}). 
For an example, LBNE10 (LBNO) can achieve CP violation discovery 
for 43\% (46\%) values of true $\dcp$ at 3$\sigma$ combining 
the expected data from the current generation experiments T2K and 
NO$\nu$A assuming $\sa$(true) = 0.5.
For 0.5*LBNO, we do not have any sensitivity at 3$\sigma$ C.L. but, 
adding the T2K and NO$\nu$A data, 37\% CP coverage can be
obtained. All these results again clearly demonstrate that the 
projected data from the current generation off-axis superbeam 
experiments will be quite useful for future generation on-axis wide 
band superbeam setups to enhance their discovery reach at higher 
confidence level.  Another important feature that emerges from 
figure~\ref{cpv-nh-lbne10-lbno10-lbno20-nova-t2k} is that the 
CP violation discovery reach is quite sensitive to the true value 
of $\sa$. The results are better if $\sa$(true) belongs to LO 
compared to HO. The main reason behind this is that like in the 
case of $\ty$~\cite{Dick:1999ed, Donini:1999jc}, the CP-asymmetry 
increases if we lower the value of $\tz$, reducing the
strength of the leading term ($C_0$) in equation~\ref{eq:matt}.

\begin{table}[tp]
\begin{center}
\begin{tabular}{|c|c|c|} \hline\hline
\multirow{2}{*}{Setups} & \multicolumn{2}{c|}{{\rule[4mm]{0mm}{4mm}Fraction of $\dcp$(true)}}
\cr\cline{2-3}
& $2\sigma$ confidence level & $3\sigma$ confidence level \cr
\hline
LBNE10 (10 kt) & 0.51 & 0.03 \cr
\hline
LBNE10 + T2K + NO$\nu$A & 0.63 & 0.43 \cr
\hline
0.5*LBNO (10 kt) & 0.40 & 0.0 \cr
\hline
0.5*LBNO + T2K + NO$\nu$A & 0.63 & 0.37 \cr
\hline
LBNO (20 kt) & 0.51 & 0.23 \cr
\hline
LBNO + T2K + NO$\nu$A & 0.69 & 0.46 \cr
\hline\hline
\end{tabular}
\caption{\footnotesize{Fraction of $\dcp$(true) for which a discovery is possible for CP violation considering NH as true hierarchy.
Here, we assume maximal mixing for the true choice of $\tz$. The results are presented at $2\sigma$ and $3\sigma$ confidence level.}}
\label{tab:CPV-compare-LBNE10-LBNO}
\end{center}
\end{table}

\section{Concluding Remarks}
\label{conclusions}

With the recent measurement of $\ty$ by reactor experiments, 
a clear and comprehensive picture of the three flavor leptonic 
mixing matrix has been established. This impressive result has 
crucial consequences for future theoretical and experimental 
efforts. It has opened up the possibility to probe the sub-dominant 
three flavor effects in both current and future long-baseline 
oscillation facilities. Another interesting piece of information 
on $\tz$ has been provided by recently completed MINOS accelerator 
experiment. $\numu \to \numu$ disappearance data of MINOS points 
towards the deviation from maximal 2-3 mixing, causing the octant 
ambiguity of $\tz$. In this paper, we present a comparative study 
of the physics reach of two future superbeam facilities, LBNE and 
LBNO in their first phases of run, in addressing the issues of 
neutrino mass hierarchy, octant of $\theta_{23}$, and leptonic CP 
violation. We also demonstrate that the projected data from current 
generation experiments, T2K and NO$\nu$A will play a crucial role 
for these future facilities to achieve their milestones with higher 
confidence level. Also for the first time, we study in detail the 
impact of the present uncertainty in 2-3 mixing angle in resolving 
these fundamental issues. 

We find that in its first phase, even a 50$\%$ scaled down version of LBNO with 10 kt detector mass has more than 7$\sigma$ 
mass hierarchy discovery reach for the lowest possible value of $\sa$(true) = 0.34 in its presently allowed 3$\sigma$ range.
This result is valid for any choices of true $\dcp$ and hierarchy. However, LBNE10 suffers in this regard and will not be able to 
provide a 5$\sigma$ result for about 50\% of the true $\dcp$ range even for maximal mixing choice for $\sa$(true).
Moreover, it fails to achieve even a 3$\sigma$ hierarchy discovery for the best-fit value in LO, $\sa$(true) = 0.41 and the worst 
case combinations of the true parameters (NH, $90^\circ$) and (IH, $-90^\circ$). In such a scenario, the projected data from
T2K and NO$\nu$A can be extremely useful for LBNE10. Adding the expected informations from T2K and NO$\nu$A,
LBNE10 can discover mass hierarchy at 3$\sigma$ confidence level for any combinations of true hierarchy-$\dcp$ and even for 
the most conservative choice of $\sa$(true) = 0.34 in its present 3$\sigma$ range. It clearly corroborates the synergy between the 
current (off-axis) and future (on-axis) superbeam experiments.

As far as the octant discovery is concerned, adding the projected data from equal neutrino and anti-neutrino runs of 
T2K (2.5 years each) and NO$\nu$A (3 years each), LBNE10 can provide a 3$\sigma$ octant resolution for $\sa$(true) $\leq$ 0.44 
and for $\sa$(true) $\geq$ 0.58 for all values of $\dcp$(true). For 0.5*LBNO, this is possible for 
$\sa$(true) $\leq$ 0.45 and for $\sa$(true) $\geq$ 0.57. 

In their first phases, both LBNE10 and LBNO can establish leptonic CP violation for around 50\% values of true $\dcp$ 
at $2\sigma$ confidence level. At $3\sigma$, their CP violation reach is quite minimal: only 3\% for LBNE10 and 23\% for LBNO. 
The expected measurements from present generation experiments T2K and NO$\nu$A can have dramatic impact on the 
CP violation discovery reach of the future facilities in their first phases of run. In case of LBNE10, CP coverage can be enhanced
from 3\% to 43\% at 3$\sigma$ combining T2K and NO$\nu$A data assuming $\sa$(true) = 0.5. For LBNO setup, CP violation discovery 
is possible for 46\% values of true $\dcp$ at 3$\sigma$ if we add the data from T2K and NO$\nu$A.



\subsubsection*{Acknowledgments}
We would like to thank M. Bishai, G. Zeller, A. Rubbia and M. Goodman for helpful communications. 
SKA was supported by the DST/INSPIRE Research Grant [IFA-PH-12], Department of Science and Technology, India.

\begin{appendix}

\section{Resolution of Octant as a function of true $\tz$ for IH(true)}
\label{appendix1}

\begin{figure}[tp]
\centering
\includegraphics[width=0.49\textwidth]{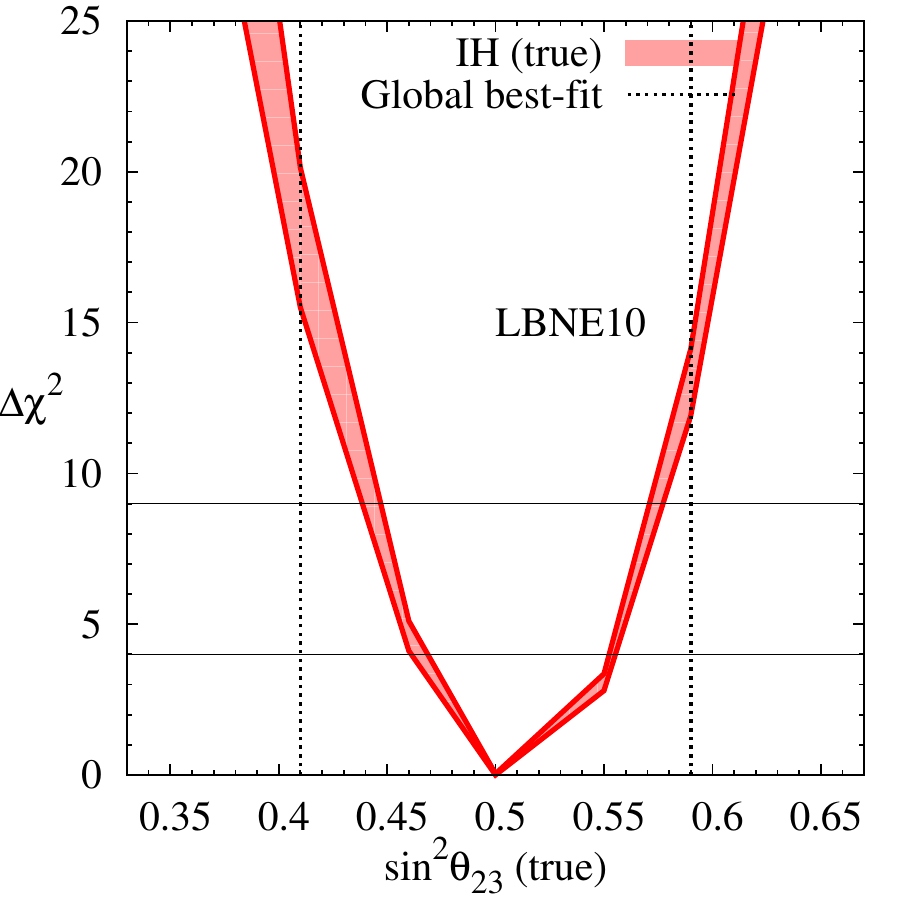}
\includegraphics[width=0.49\textwidth]{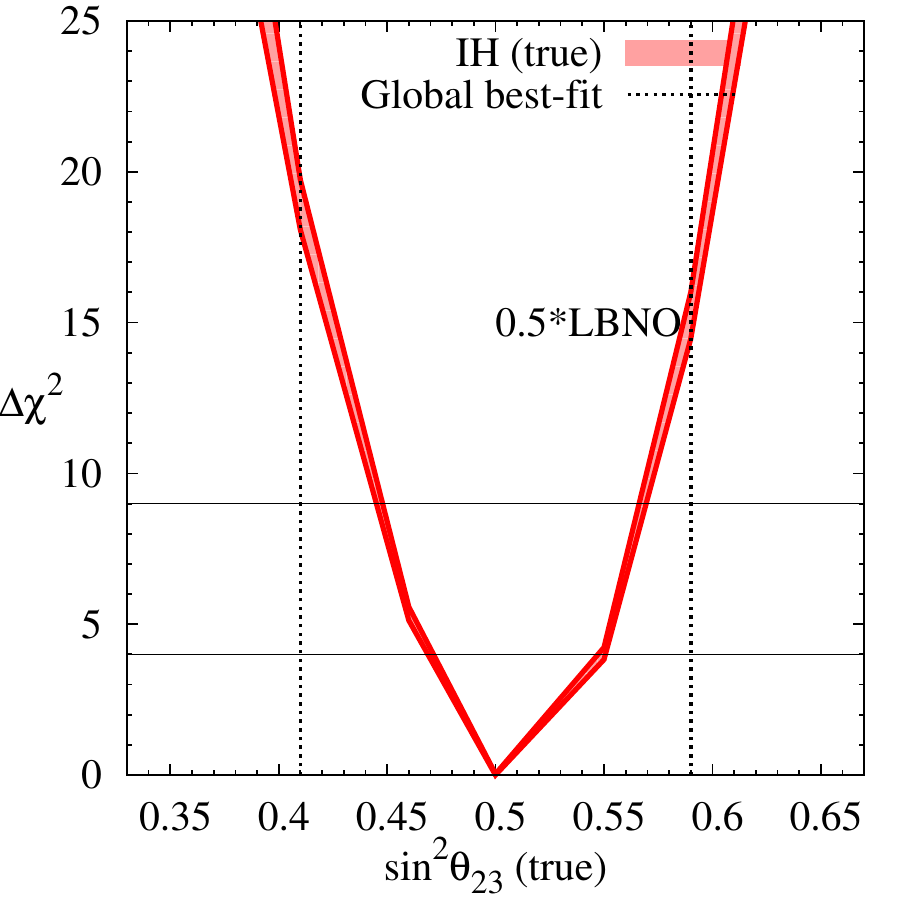}
\caption{\footnotesize{$\dchsq_{{\footnotesize \textrm{min}}}$ for octant resolution as a function of true $\sa$. Left panel (right panel) is for 
LBNE10 (0.5*LBNO). The variation due to $\dcp$(true) leads to the band in $\dchsq$ for a given $\sa$(true).
The vertical lines correspond to the global best-fit values. We consider IH as true hierarchy. In producing all these plots, 
the projected data from T2K and NO$\nu$A have been added (see section~\ref{experiments} for details).}}
\label{dchsq_vs_true_tz_ih_true}
\end{figure}

In this appendix, we present the results for octant discovery generating the data with IH. In figure~\ref{dchsq_vs_true_tz_ih_true}, 
we show the $\dchsq_{{\footnotesize \textrm{min}}}$ as a function of true $\sa$ for LBNE10 (left panel) and 0.5*LBNO (right panel) 
assuming IH as true hierarchy. Variation of $\dcp$(true) in the range $-180^\circ$ to $180^\circ$ leads to the band in $\dchsq$ 
values for a given true $\sa$. The vertical lines indicate towards the global best-fit values. Here we add the projected data from T2K 
and NO$\nu$A to produce these results. For LBNE10, a 3$\sigma$ octant resolution is possible for 
$\sa$(true) $\leq$ 0.44 and for $\sa$(true) $\geq$ 0.58 irrespective of the values of $\dcp$(true). For 0.5*LBNO, this is possible for 
$\sa$(true) $\leq$ 0.44 and for $\sa$(true) $\geq$ 0.57. We see that the results with IH(true) choice are quite similar to that of NH(true)
(see figure~\ref{dchsq_vs_true_tz_nh_true}).

\begin{figure}[tp]
\centering
\includegraphics[width=0.49\textwidth]{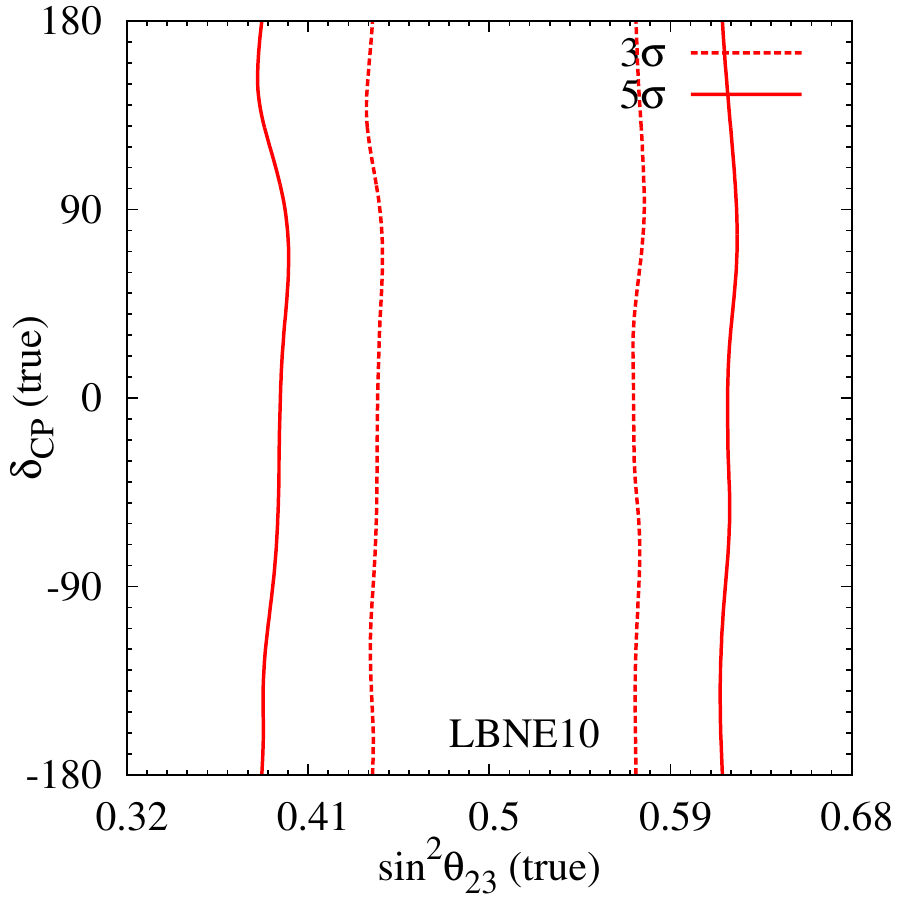}
\includegraphics[width=0.49\textwidth]{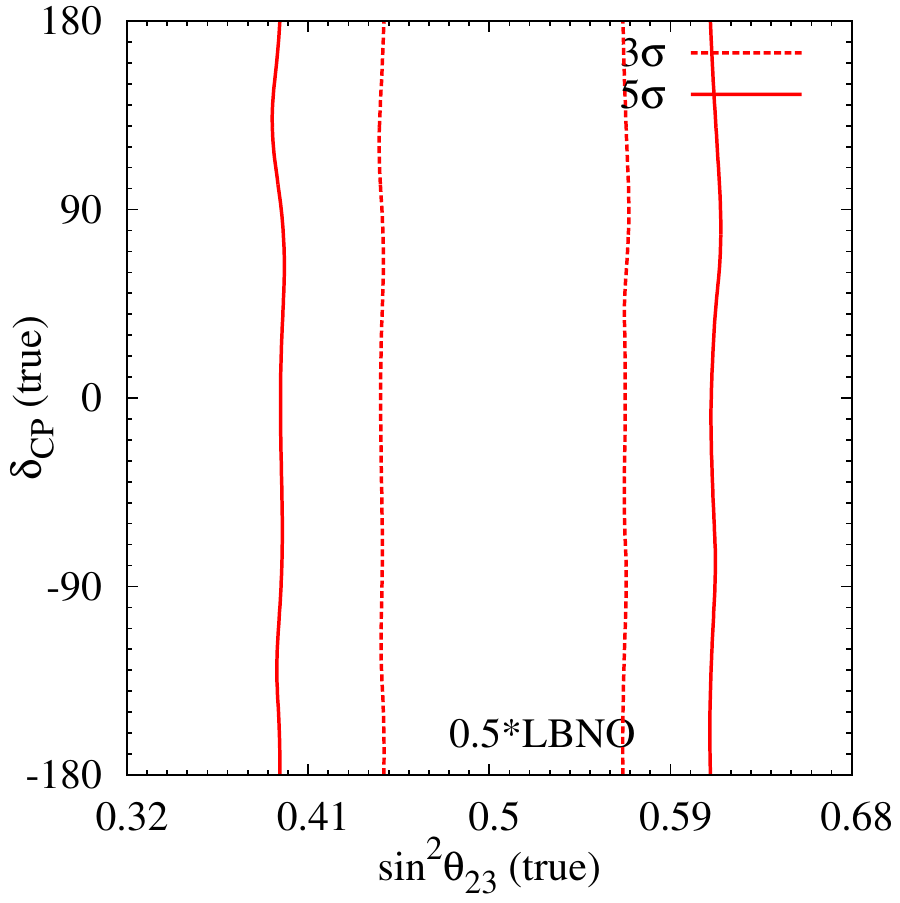}
\caption{\footnotesize{Octant resolving capability at 3$\sigma$ and 5$\sigma$ C.L. in the true $\sa$ - true $\dcp$ 
plane for LBNE10 (left panel) and 0.5*LBNO (right panel). The vertical lines point towards the global best-fit values. 
Here, we assume IH as true hierarchy. In generating all these plots, the projected data from 
T2K and NO$\nu$A have been added (see section~\ref{experiments} for details).}}
\label{true-tz-true-cp-ih-true}
\end{figure}

Figure~\ref{true-tz-true-cp-ih-true} shows the 3$\sigma$ and 5$\sigma$ octant resolution contours in 
true $\sa$ - true $\dcp$ plane considering IH as true hierarchy. The left (right) panel is for LBNE10 (0.5*LBNO) 
adding the expected data from T2K and NO$\nu$A. Octant resolution is only possible for points lying outside the contours. 
This figure again suggests that for IH(true) case, both LBNE10 and 0.5*LBNO in combination with T2K and NO$\nu$A data 
can provide octant discovery for global best-fit points at 3$\sigma$ confidence level.

\section{CP Violation discovery as a function of true $\dcp$ for IH(true)}
\label{appendix2}

\begin{figure}[tp]
\centering
\includegraphics[width=0.325\textwidth]{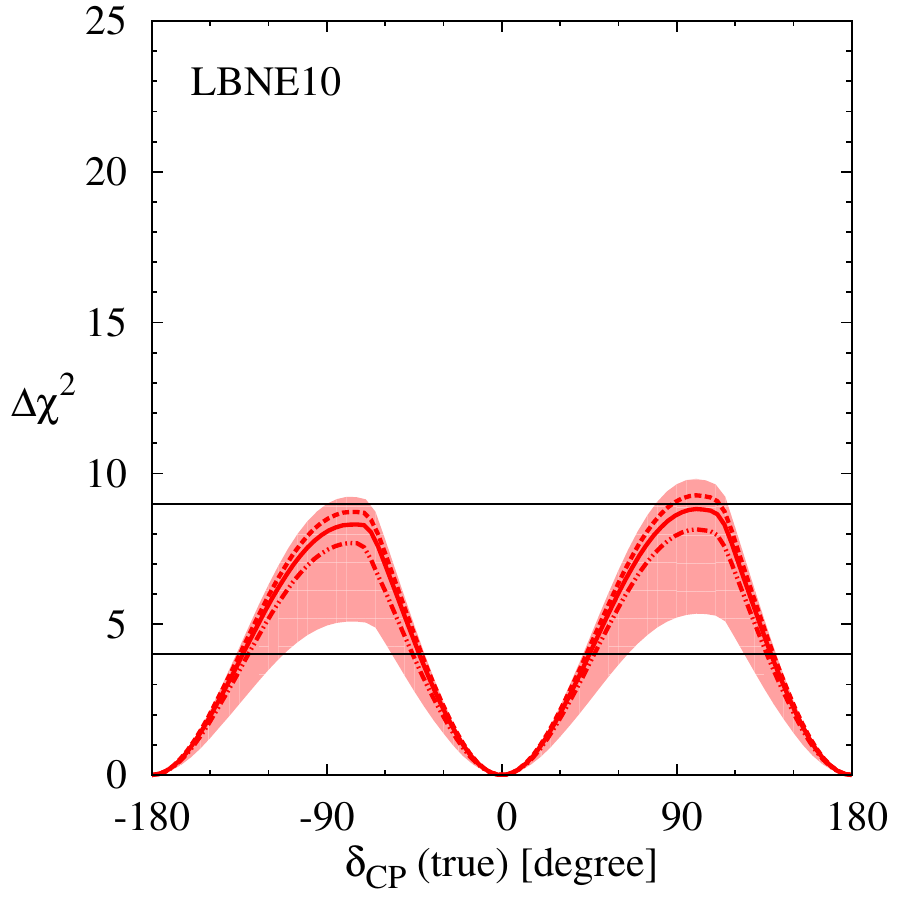}
\includegraphics[width=0.325\textwidth]{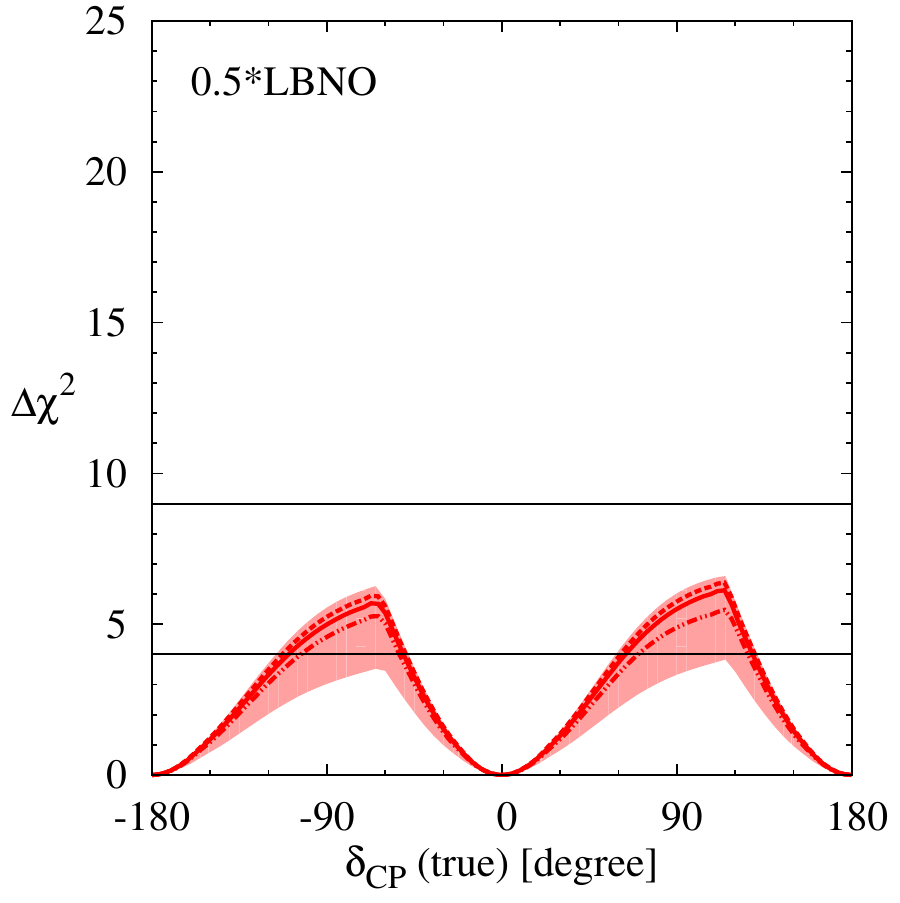}
\includegraphics[width=0.325\textwidth]{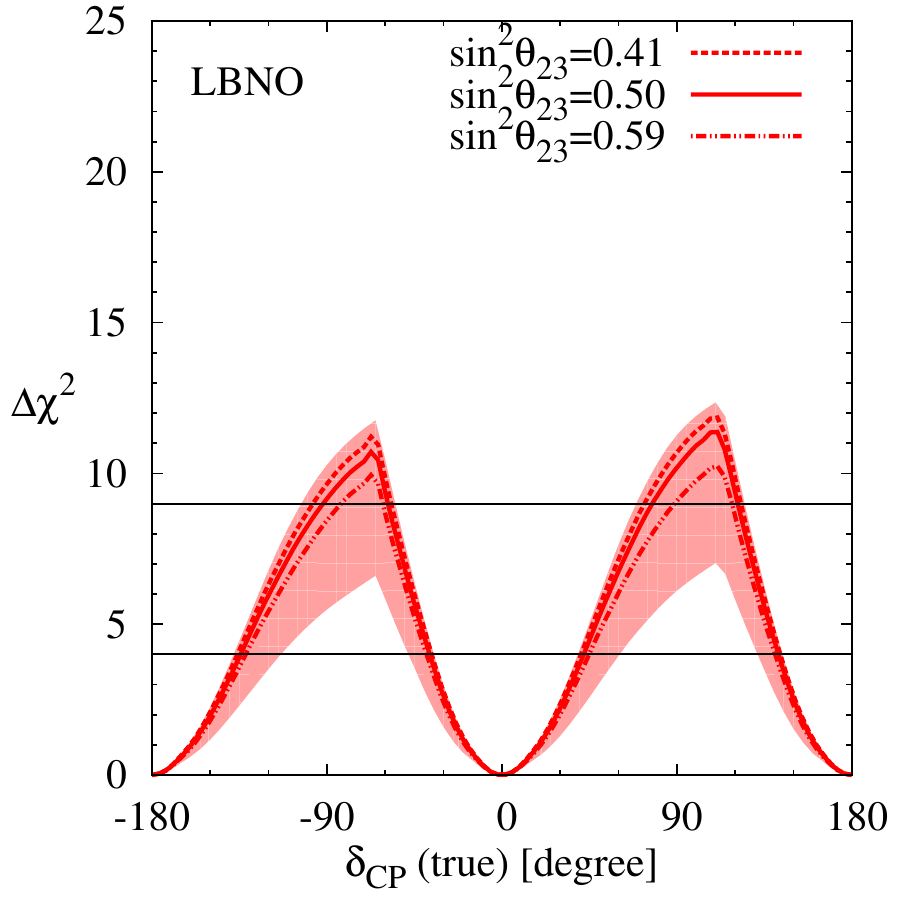}
\vskip0.5cm
\includegraphics[width=0.325\textwidth]{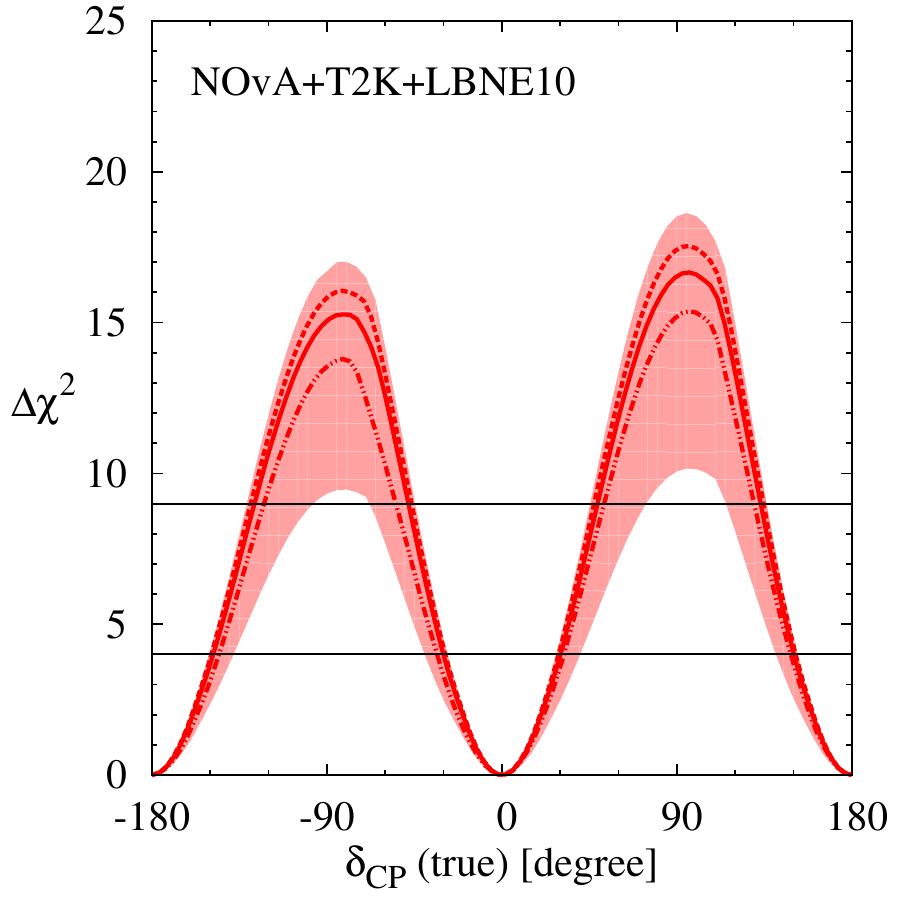}
\includegraphics[width=0.325\textwidth]{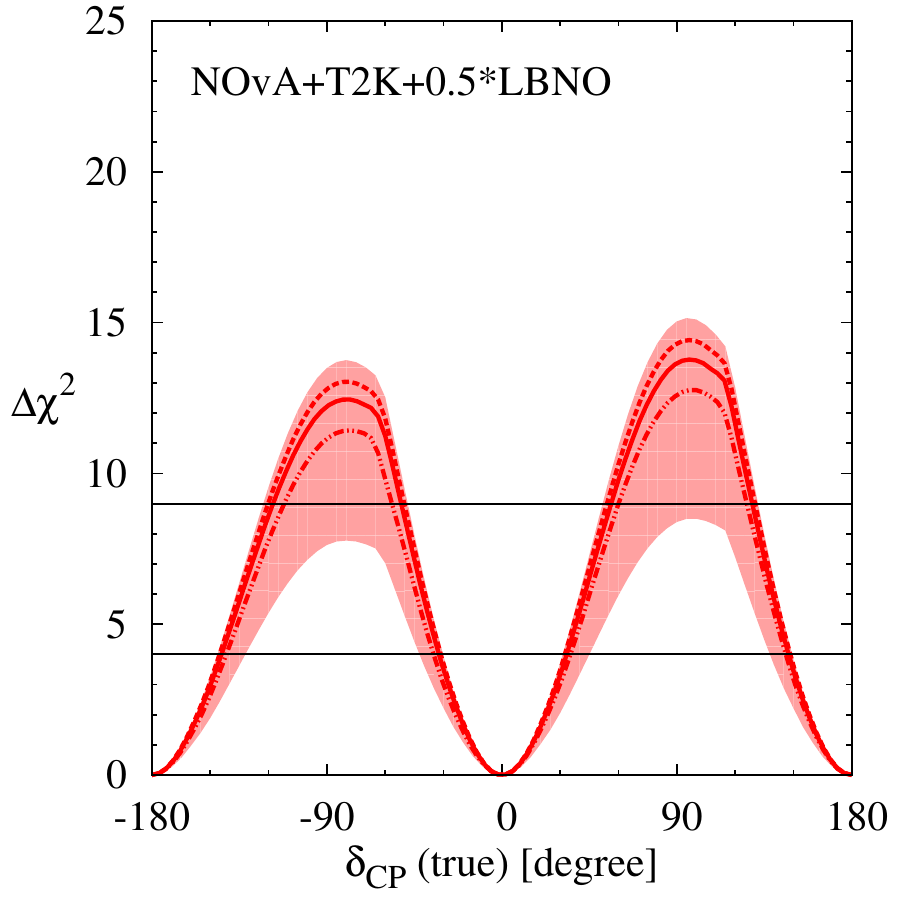}
\includegraphics[width=0.325\textwidth]{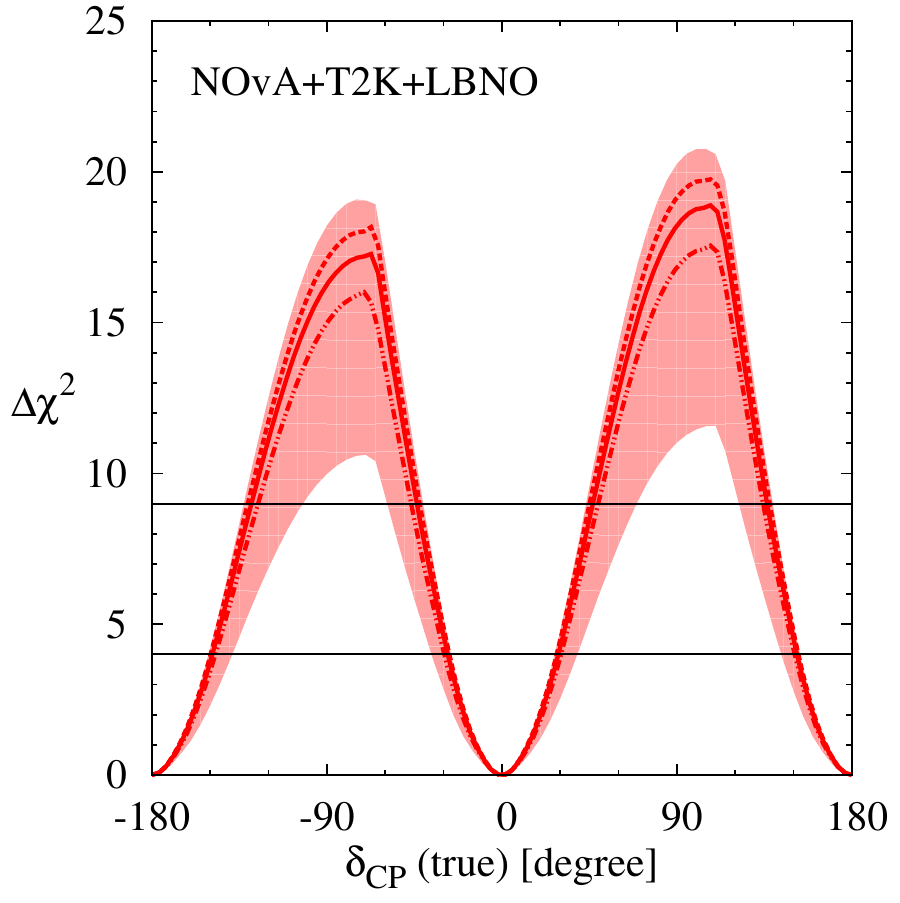}
\caption{\footnotesize{CP Violation discovery reach as a function of true value of $\dcp$ assuming IH as true hierarchy. 
Results are shown for LBNE10 (10 kt), 0.5*LBNO (10 kt), and LBNO (20 kt) in the left, middle, and right upper panels respectively. 
In lower panels, we depict the same adding the information from T2K and NO$\nu$A experiments. The shaded band depicts
the variation in $\dchsq_{{\footnotesize \textrm{min}}}$ due to different true choices of $\sa$ in its 3$\sigma$ allowed range of 
0.34 to 0.67. Inside the band, we show the results for three different true values of $\sa$: 0.41, 0.5, and 0.59.}}
\label{cpv-ih-lbne10-lbno10-lbno20-nova-t2k}
\end{figure}

In figure~\ref{cpv-ih-lbne10-lbno10-lbno20-nova-t2k}, we give the CP violation discovery reach for various experimental setups 
under study as a function of true $\dcp$ considering IH as true hierarchy. Like in figure~\ref{cpv-nh-lbne10-lbno10-lbno20-nova-t2k},  
the left, middle, and right upper panels of figure~\ref{cpv-ih-lbne10-lbno10-lbno20-nova-t2k} present the results for 
LBNE10, 0.5*LBNO, and LBNO respectively. In lower panels, we depict the same combining the projected data from 
T2K and NO$\nu$A experiments. The shaded band in each panel reflects the variation in $\dchsq_{{\footnotesize \textrm{min}}}$ due 
to different true choices of $\sa$ in its 3$\sigma$ allowed range of 0.34 to 0.67. Inside the band, we give the results for three different 
true values of $\sa$: 0.41, 0.5, and 0.59. We do not see any qualitative differences in the CP violation discovery reach for these setups
when we generate the data assuming IH instead of NH. Around $\dcp$(true) = $\pm$ $90^\circ$ (where CP violation is maximum), 
the results are slightly better for NH(true) compared to IH(true).

\end{appendix}

\bibliographystyle{apsrev}
\bibliography{ical-references}

\end{document}